\documentclass[review]{elsarticle}

\usepackage{lineno,hyperref}
\usepackage{amsmath,amsfonts,amsthm}
\modulolinenumbers[5]

\newcommand{\N}{{\mathbb{N}}}  
\newcommand{\Z}{{\mathbb{Z}}}  
\newcommand{\Q}{{\mathbb{Q}}} 
\newcommand{\A}{{\mathbb{A_R}}}  
\newcommand{\C}{{\mathbb{C}}}  
\newcommand{\F}{{\mathbb{F}}}  
\newcommand{\0}{\mathbf{0}} 
\newcommand{\qfa}{\mathcal{Q}} 
\newcommand{\Gen}{{\mathcal{G}}}  
\newcommand{\Genn}{\Gamma}  
\newcommand{\HQ}{{\mathbb{H}}}  

\newcommand{\ov}[1]{\overline{#1}}
\newcommand\abs[1]{\left\vert #1\right\vert}

\newcommand{\norm}[1]{\left|\left|{#1}\right|\right|}

\theoremstyle{plain}
\newtheorem{lemma}{Lemma}

\newtheorem{prob}{Problem}
\newtheorem{definition}{Definition}
\newtheorem{example}{Example}
\newtheorem{open}{Open Problem}
\newtheorem{theorem}{Theorem}
\newtheorem{remark}{Remark}
\newtheorem{corollary}{Corollary}

\begin{document}

\begin{frontmatter}

\title{On Injectivity of Quantum Finite Automata}
\tnotetext[mytitlenote]{Published version of this article available at: https://doi.org/10.1016/j.jcss.2021.05.002}

\author{Paul C. Bell}
\address{Department of Computer Science, Byrom Street, Liverpool~John~Moores~University, Liverpool, L3-3AF, UK, p.c.bell@ljmu.ac.uk, https://orcid.org/0000-0003-2620-635X}

\author{Mika Hirvensalo}
\address{Department of Mathematics and Statistics, University of Turku, FI-20014, Turku, Finland, mikhirve@utu.fi}

\begin{keyword}
Quantum finite automata; matrix freeness; undecidability; Post's correspondence problem, quaternions.
\end{keyword}

\begin{abstract} We consider notions of freeness and ambiguity for the acceptance probability of Moore-Crutchfield Measure Once Quantum Finite Automata (MO-QFA). We study the \emph{injectivity} problem of determining if the acceptance probability function of a MO-QFA is injective over all input words, i.e., giving a distinct probability for each input word. We show that the injectivity problem is undecidable for $8$ state MO-QFA, even when all unitary matrices and the projection matrix are rational and the initial state vector is real algebraic. We also show undecidability of this problem when the initial vector is rational, although with a huge increase in the number of states. We utilize properties of quaternions, free rotation groups, representations of tuples of rationals as linear sums of radicals and a reduction of the mixed modification of Post's correspondence problem, as well as a new result on rational polynomial packing functions which may be of independent interest.
\end{abstract}

\end{frontmatter}


\section{Introduction}
Measure-Once Quantum Finite Automata (QFA) were introduced in \cite{MC} as a natural quantum variant of probabilistic finite automata. The model is defined formally in Section~\ref{sec-qfa}, but briefly a QFA over an alphabet $\Sigma$ is defined by a three tuple $\mathcal{Q} = (P, \{X_a | a \in \Sigma\}, u)$ where $P$ is a projection matrix, $X_a$ is a complex unitary matrix for each alphabet letter $a \in \Sigma$ and $u$ is a unit length vector with respect to the Euclidean ($\ell^2$) norm. Given an input word $w = w_1 \cdots w_k \in \Sigma^*$, then the acceptance probability $f_{\mathcal{Q}}: \Sigma^* \to \mathbb{R}$ of $w$ under $\mathcal{Q}$ is given by 
$$f_{\mathcal{Q}}(w)=\norm{PX_{w_k} \cdots X_{w_1}u}^2.$$ 

We also denote the acceptance probability by $\mathcal{Q}(w)$ with a slight abuse of notation.  The related model of Probabilistic Finite Automata (PFA) with $n$ states over an alphabet $\Sigma$ is defined as $\mathcal{P}=(x, \{M_a|a \in \Sigma\}, y)$ where $y \in \mathbb{R}^n$ is the initial probability distribution (unit length under $\ell^1$ norm); $x \in \{0, 1\}^n$ is the final state vector and each $M_a \in \mathbb{R}^{n \times n}$ is a stochastic matrix. For a word $w = w_1\cdots w_k \in \Sigma^*$, we define the acceptance probability $f_{\mathcal{P}}: \Sigma^* \to \mathbb{R}$ of $\mathcal{P}$ as:
$$
f_{\mathcal{P}}(w) = x^TM_{w_k} \cdots M_{w_1} y.
$$

For any $\lambda \in [0, 1]$ and automaton $\mathcal{A}$ (either PFA or QFA) over alphabet $\Sigma$, we define a cut-point language to be: $L_{\geq \lambda}(\mathcal{A}) = \{w \in \Sigma^* | f_{\mathcal{A}}(w) \geq \lambda\}$, and a strict cut-point language $L_{> \lambda}(\mathcal{A})$ by replacing $\geq$ with $>$. The (strict) emptiness problem for a cut-point language is to determine if $L_{\geq \lambda}(\mathcal{A}) = \emptyset$ (resp. $L_{> \lambda}(\mathcal{A}) = \emptyset$). 

The QFA model is restricted due to unitarity constraints and can recognize only group languages (those regular languages whose syntactic monoid is a group \cite{Brodsky2002}). Whereas the emptiness problem for strict cut-point languages is undecidable for PFA \cite{Paz}, it surprisingly becomes decidable for QFA \cite{BJK05}. The decidability is established via the compactness of the group generated by unitary matrices: a compact algebraic group has a finite polynomial basis, and the decision procedure is then based on Tarski's quantifier elimination theorem \cite{BJK05}.

Another surprising result was also presented in \cite{BJK05}: namely that the emptiness problem for non-strict (allowing equality) cut-point languages on QFA is undecidable. The sizes of the automata exhibiting undecidability were subsequently improved in \cite{Hir}.
As these examples illuminate, the border between decidability and undecidability may be crossed with a minor modification to the model or premises.

Reachability problems for matrix semigroups have attracted a great deal of attention over the past few years. Typically, we are given a finite set of generating matrices $\mathcal{G}$ forming a semigroup $\mathcal{S} = \langle \mathcal{G} \rangle$ and we ask some question about $\mathcal{S}$. A seminal result of Paterson showed that the mortality problem for integer matrix semigroups is undecidable in dimension three \cite{Paterson}. In this problem, $\mathcal{G} \subseteq \mathbb{Z}^{3 \times 3}$ and we ask whether the zero matrix belongs to $\mathcal{S} = \langle \mathcal{G} \rangle$. Determining if the identity matrix belongs to the semigroup generated by a given set of generating matrices was later shown to be undecidable for four-dimensional integer matrices \cite{BP10}.

A related problem is the freeness problem for integer matrices --- given a set $\mathcal{G} \subseteq \mathbb{F}^{n \times n}$, where $\mathbb{F}$ is a semiring,  determine if $\mathcal{G}$ is a code for the semigroup generated by $\mathcal{G}$ (i.e., if every element of $\langle \mathcal{G} \rangle$ has a unique factorization over elements of $\mathcal{G}$). It was proven by Klarner et al. that the freeness problem is undecidable over $\mathbb{N}^{3 \times 3}$ in \cite{KBS91} and this result was improved by Cassaigne et al. to hold even for \emph{upper-triangular} matrices over $\mathbb{N}^{3 \times 3}$ in \cite{CHK99}. 

There are many open problems related to freeness in $2 \times 2$ matrices; see \cite{CN12, CH14, CK05} for good surveys. The freeness problem over $\mathbb{H}^{2 \times 2}$ is undecidable \cite{BP08quat}, where $\mathbb{H}$ is the skew-field of quaternions (in fact the result even holds when all entries of the quaternions are rationals). 

The freeness problem for matrix equations of a specific form was recently studied. Given a finite set of matrices $\{M_1, \ldots, M_k\} \subseteq \mathbb{Q}^{n \times n}$, we may consider a set of matrices of the form: $\{M_1^{j_1} \cdots M_k^{j_k} | j_i \geq 0 \textnormal{ where } 1 \leq i \leq k\}$. The freeness problem for such a set asks if there exists a choice of variables such that $j_1, \ldots, j_k, j_1', \ldots, j_k' \geq 0$, where at least one $j_i \neq j_i'$ such that $M_1^{j_1} \cdots M_k^{j_k} = M_1^{j_1'} \cdots M_k^{j_k'}$ in which case the set of matrices is not free. This problem was shown to be decidable when $n = 2$, but undecidable in general \cite{CH14}.

In a similar vein, we may study the vector freeness and ambiguity problems, where we are given a finite set of matrices $\mathcal{G} \subseteq{\mathbb{F}}^{n \times n}$ and a vector $u \in \mathbb{F}^{n}$. The \emph{ambiguity problem} is to determine whether there exist two matrices $M_1, M_2 \in \mathcal{S} = \langle \mathcal{G} \rangle$ with $M_1 \neq M_2$ such that $M_1u = M_2u$ and the \emph{freeness  problem} is to determine the uniqueness of factorizations of $\{Mu | M \in \mathcal{S}\}$ i.e., does $M_{i_1} \cdots M_{i_k}u =  M_{j_1} \cdots M_{j_{k'}}u$, where each $M_t \in \mathcal{G}$, imply that $k = k'$ and $M_{i_r} = M_{j_r}$ for $1 \leq r \leq k$? The difference between the problems is thus whether we search for two distinct matrices in the semigroup $\mathcal{S}$, or else two distinct factorizations of matrices from $\Gen$ (possibly leading to the same matrix) which, when multiplied by $u$, lead to the same vector. 

It should be noted that these (related but distinct) problems seem more difficult to solve than freeness for matrix semigroups, since by multiplying matrix $M_1$ and $M_2$ with $u$, some information may be lost. A motivation for such a problem is that many linear dynamic systems/computational models are defined in this way. The freeness question now asks whether starting from some initial point, we have two separate computational paths which coincide at some later point, or else whether every configuration starting from $u$ is uniquely determined. These questions were studied in \cite{BPSofsem08} and the problems were shown to be undecidable when $\mathcal{S} \subseteq \mathbb{Z}^{4 \times 4}$, or when $\mathcal{S} \subseteq \mathbb{Q}^{3 \times 3}$. The NP-completeness of the vector ambiguity and freeness problems for SL($2, \mathbb{Z}$) was recently shown\footnote{In fact, the freeness problem, as it is usually formulated, is coNP-complete; NP-completeness refers to the `non-freeness' counterpart of this problem.} in \cite{KP18} (where SL($2, \mathbb{Z}$) is the special linear group of $2 \times 2$ matrices). 

Whilst vector reachability questions are interesting from the point of view of dynamical systems, many computational models have the notion of a projection being taken at the end. This usually takes the form of defining a partition of configurations into accepting or nonaccepting states. This leads to the notion of \emph{scalar reachability} (also known as half-space reachability \cite{COSW}), defined in terms of \emph{two} vectors, $u$ and $v$, where we now study the set of scalars  $\{u^TMv | M \in \mathcal{S}\}$. The \emph{scalar ambiguity} question then asks whether or not this set of scalars is unique i.e., does there exist two matrices $M_1, M_2 \in \mathcal{S}$ with $M_1 \neq M_2$ such that $u^TM_1v = u^TM_2v$? The difficulty  with extending the undecidability result for vector reachability is that all information about each matrix $M$ needs to be stored within a single scalar value $u^TMv$ in a unique way.

In \cite{BCJ16}, the freeness or \emph{injectivity} problem (defined formally in Section~\ref{scalDef}) for $4$-state weighted and $6$-state probabilistic automata was shown to be undecidable. Essentially, the injectivity problem asks if the acceptance probabilities of all input words are distinct. The undecidability result was shown to hold even when the input words come from a bounded language, thus the matrices appear in some fixed order, and are taken to an arbitrary power.  The problem can also be stated in terms of \emph{formal power series}: given a formal power series $r$, determine if $r$  has two equal coefficients, as studied in \cite{KS86} and Theorem~3.4 of \cite{Ho98}. As mentioned above, several reachability problems for PFA (such as emptiness of cut-point languages) are known to be \emph{undecidable} \cite{Paz}, even in a fixed dimension \cite{BC03, Hir}. The reachability problem for PFA defined on a bounded language (i.e., where input words are from a bounded language which is given as part of the input) was shown to be undecidable in \cite{BHH13}. We may note that the scalar freeness and ambiguity problems are a similar concept to the \emph{threshold isolation problem} which asks whether a given cutpoint may be approached arbitrarily closely and which is known to be undecidable \cite{BMT77, BC03}.

It is therefore natural to ask whether the {\em injectivity problem} is undecidable for QFA. This problem appears more difficult to prove than for weighted or probabilistic automata, since the acceptance probability of a QFA $\mathcal{Q}$ has the form $\mathcal{Q}(w) = \norm{PX^Ru}^2$ and it is thus difficult to encode sufficient information about the matrix $X$ within $\mathcal{Q}(w)$ to guarantee uniqueness of the acceptance probabilities. We show that the injectivity and ambiguity problems are undecidable for $8$ (resp. $9$) state QFA by using an encoding of the mixed modification of Post's Correspondence Problem and a result related to linear independence of rationals of a basis of squarefree radicals as well as  techniques from linear algebra and properties of quaternions. We then show that the problem remains undecidable even if the initial vector is rational, albeit with a huge increase in the number of states. To achieve this aim, we prove a new result on packing functions of $n$-tuples of rationals of a specific form in Section~\ref{nonrad}, which may be of independent interest.

A preliminary version of this paper appeared in \cite{BH19}. The present paper significantly reduces the number of states required for showing the undecidability of the injectivity problem for QFA from $32$ to $8$ and extends the result to rational QFA (i.e. QFA where all components are over the rationals rather than algebraic reals).

\section{Notation}\label{sec-not}

Let $\Sigma = \{x_1, x_2, \ldots,x_k\}$ be a finite set of \emph{letters} called an \emph{alphabet}. A word $w$ is a finite sequence of letters from $\Sigma$, the set of all words over $\Sigma$ is denoted $\Sigma^*$ and the set of nonempty words is denoted $\Sigma^+$. The \emph{empty word} is denoted by $\varepsilon$. We use $|u|$ to denote the length of a word $u$ and thus $|\varepsilon|=0$. For two words $u = u_1u_2 \cdots u_i$ and $v = v_1v_2 \cdots v_j$, where $u,v \in \Sigma^*$, the concatenation of $u$ and $v$ is denoted by $u\cdot v$ (or by $uv$ for brevity) such that $u\cdot v = u_1u_2 \cdots u_iv_1v_2 \cdots v_j$. Word $u^R = u_{i} \cdots u_2u_1$ denotes the mirror image or reverse of word $u$. For a word $u$, we denote by $u^*$ zero or more concatenations of  $u$ with itself, i.e., $u^* = \{\epsilon, u, uu, \ldots\}$. A subset $L$ of $\Sigma^*$ is called a \emph{language}. A language $L \subseteq \Sigma^*$ is called a \emph{bounded language} if and only if there exist words $w_1, w_2, \ldots, w_m \in \Sigma^+$ such that $L \subseteq w_1^* w_2^* \cdots w_m^*$.  Given an alphabet $\Sigma$ as above, we denote by $\Sigma^{-1}$  the set $\{x_1^{-1}, \ldots, x_k^{-1}\}$, where each $x_i^{-1}$ is a new letter with the property that $x_ix_i^{-1} = x_i^{-1}x_i = \varepsilon$ are the only identities of the group $\langle \Sigma \rangle_{\textrm{gp}}$ (the group generated by $\Sigma \cup \Sigma^{-1}$). A word $w = w_1 w_2 \cdots w_i \in (\Sigma \cup \Sigma^{-1})^*$ is called \emph{reduced} if there does not exist $1 \leq j < i$ such that $w_jw_{j+1} = \varepsilon$; i.e., no two consecutive letters are inverse. 

Given any two rings $R_1$ and $R_2$ we use the notation $R_1 \hookrightarrow R_2$ to denote a \emph{monomorphism} i.e., an injective homomorphism between $R_1$ and $R_2$. Given a finite set $\mathcal{G}$, we use the notation $\langle \mathcal{G} \rangle$ (resp. $\langle \mathcal{G} \rangle_{\textrm{gp}}$) to denote the \emph{semigroup} (resp. group) generated by $\mathcal{G}$. 

\subsection*{Semirings and quaternions}

We denote by $\N$ the natural numbers, $\Z$ the integers, $\Q$ the rationals, $\C$ the complex numbers and $\HQ$ the quaternions. We denote by $\C(\Q)$ the complex numbers with rational parts, by $\HQ(\Q)$ the quaternions with rational parts and by $\A$ the real algebraic numbers. 

Given any semiring $\F$ we denote by $\F^{i \times j}$ the set of $i \times j$ matrices over $\F$. We denote by $e_i$ the $i$'th basis vector of some dimension (which will be clear from the context or else explicitly stated).

In a similar style to complex numbers, a rational quaternion $\vartheta \in \HQ(\Q)$ can be written $\vartheta = a + b\mathbf{i} + c\mathbf{j} + d\mathbf{k}$ where $a,b,c,d \in \mathbb{Q}$. 
To ease notation let us define the vector:
$\mu = (1, \mathbf{i}, \mathbf{j}, \mathbf{k})$ and it is now clear that 
$\vartheta = (a,b,c,d) \cdot \mu$ where $\cdot$ denotes the inner or `dot' product.

Quaternion addition is simply the componentwise addition of elements.
It is well known that quaternion multiplication is not commutative (hence they form a skew field). Multiplication is completely defined by the equations 
$\mathbf{i}^2 = \mathbf{j}^2 = \mathbf{k}^2 = -1$ , $\mathbf{ij} = \mathbf{k} = -\mathbf{ji}$, $\mathbf{jk} = \mathbf{i} = -\mathbf{kj}$ and 
$\mathbf{ki} = \mathbf{j} = -\mathbf{ki}$. Thus for two quaternions $\vartheta_1 = (a_1, b_1, c_1, d_1)\mu$ and $\vartheta_2 = (a_2, b_2, c_2, d_2) \mu$,
we can define their product as
$
\vartheta_1 \vartheta_2 =  (a_1 a_2 - b_1b_2 - c_1c_2 - d_1d_2)
  + (a_1b_2 + b_1a_2 + c_1d_2 - d_1c_2) \mathbf{i} 
   + (a_1c_2 - b_1d_2 + c_1a_2 + d_1b_2) \mathbf{j}
   + (a_1d_2 + b_1c_2 - c_1b_2 + d_1a_2) \mathbf{k}.
$

In a similar way to complex numbers, we define the conjugate of $\vartheta = (a,b,c,d)\cdot \mu$ by 
$\overline{\vartheta} = (a,-b,-c,-d)\cdot \mu$.
We can now define a norm on the quaternions by $|| \vartheta || = \sqrt{ \vartheta \overline{ \vartheta}} = \sqrt{a^2 + b^2 + c^2 + d^2}$.
Any non zero quaternion has a multiplicative (and obviously an additive) inverse \cite{Le04}. 
The other properties of being a skew field can be easily checked.

A \emph{unit} quaternion (norm $1$) corresponds to a rotation in three dimensional space \cite{Le04}.

\subsection*{Linear Algebra}

Given $A = (a_{ij}) \in\F^{m\times m}$ and $B\in\F^{n\times n},$ we define the direct sum $A\oplus B$ and Kronecker product $A \otimes B$ of $A$ and $B$ by:
\[
A\oplus B=
\left[\begin{array}{@{}c|l@{}}
A & \0_{m,n}\\
\hline
\0_{n,m} & B
\end{array}\right], \quad
A\otimes B=
\left[\begin{array}{cccc}
a_{11}B & a_{12}B & \cdots & a_{1m}B \\ 
a_{21}B & a_{22}B & \cdots & a_{2m}B \\ 
\vdots & \vdots & \ddots &\vdots \\ 
a_{m1}B & a_{m2}B & \cdots & a_{mm}B \\ 
\end{array}\right],
\]
where $\0_{i,j}$ denotes the zero matrix of dimension $i \times j$. Note that neither $\oplus$ nor $\otimes$ are commutative in general. For two vectors $u = (u_1, \ldots, u_m)^T \in \mathbb{F}^m$ and $v = (v_1, \ldots v_n)^T  \in \mathbb{F}^n$ then we define $u \oplus v = (u_1, \ldots, u_m, v_1, \ldots, v_n)^T \in \mathbb{F}^{m+n}$ by a minor abuse of notation. Given a finite set of matrices $\mathcal{G} = \{G_1, G_2, \ldots, G_m\} \subseteq \mathbb{F}^{n \times n}$, $\langle\mathcal{G}\rangle$ is the semigroup generated by $\mathcal{G}$. We will use the following notations:
$$
\bigoplus_{j=1}^{\ell} G_j = G_1 \oplus G_2 \oplus \cdots \oplus G_\ell, \qquad
\bigotimes_{j=1}^{\ell} G_j = G_1 \otimes G_2 \otimes \cdots \otimes G_\ell.
$$

Given a single matrix $G \in \mathbb{F}^{n \times n}$, we inductively define $G^{\otimes k} = G \otimes G^{\otimes (k-1)} \in \mathbb{F}^{n^k \times n^k}$  with $G^{\otimes 1} = G$ as the $k$-fold Kronecker power of $G$ for $k \geq 1$. Similarly,  $G^{\oplus k} = G \oplus G^{\oplus (k-1)} \in \mathbb{F}^{nk \times nk}$  with $G^{\oplus 1} = G$. The following properties of $\oplus$ and $\otimes$ are well known; see \cite{HJ91} for proofs.

\begin{lemma}\label{kronprop}
Let $A, B, C, D \in \mathbb{F}^{n \times n}$. Then:
\begin{itemize}
\item $(A \otimes B) \otimes C = A \otimes (B \otimes C)$ and $(A \oplus B) \oplus C = A \oplus (B \oplus C)$, thus $A \otimes B \otimes C$ and $A \oplus B \oplus C$ are unambiguous.
\item Mixed product properties: $(A \otimes B)(C \otimes D) = (AC) \otimes (BD)$ and $(A \oplus B)(C \oplus D) = (AC) \oplus (BD)$.
\item If $A$ and $B$ are unitary matrices, then so are $A \oplus B$ and $A \otimes B$. 
\end{itemize}
\end{lemma}

We now move to formally define quantum automata and the mixed modification Post's correspondence problem which will later be used in our reductions.

\section{Quantum Finite Automata and Undecidability}\label{sec-qfa}

Our main computational model of interest will be (measure-once) quantum finite automata, which we now define. 

\begin{definition}\label{MOQFA}
A measure-once $n$-state quantum automaton (QFA) over a $k$-letter alphabet $\Sigma$ is a triplet $(P,\{X_a\mid a\in\Sigma\},u)$, where $P\in{\mathbb C}^{n\times n}$ is a projection matrix, each
$X_a\in{\mathbb C}^{n\times n}$ is a unitary matrix (where rows form an orthonormal set), and $u\in{\mathbb C}^{n}$ is a unit-length vector.

A morphism $\Sigma^*\to\langle X_a\rangle$ is defined as
$w=a_{i_1}\ldots a_{i_t}\mapsto X_w\stackrel{\text{def}}{=}X_{a_{i_1}}\ldots X_{a_{i_t}}$  and the {\em acceptance probability function} of a QFA $\mathcal{Q}$ is defined as $\mathcal{Q}(w)=\norm{PX_{w^R}u}^2$ by abuse of notation. We use the reverse of the word $w$, denoted $w^R$, so that $w_1$ is applied first, then $w_2$ etc. 
\end{definition}

\subsection{Ambiguity and Injectivity for QFA}\label{scalDef}

Consider a finite set of unitary matrices $\Gen = \{X_1, X_2, \ldots, X_k\} \subset\C^{n\times n}$, a projection matrix $P \in \C^{n \times n}$ and a unit-length (according to the $l^2$-norm)  column vector $u \in\C^n$. Let $\qfa = (P, \Gen, u)$ be a QFA and define $\Lambda(\qfa)$ to be the set of scalars such that $\Lambda(\qfa)=\{\left|\left|PXu\right|\right|^2 ; X \in \langle\Gen\rangle\}$. 

If the acceptance function $\qfa(w)$ is injective, then the QFA is said to be injective. In other words, a QFA $\qfa$ is injective if $\qfa(w_1) = \qfa(w_2)$ for $w_1, w_2 \in \Sigma^*$ implies $w_1 = w_2$.

If for $\lambda \in \Lambda(\qfa)$ there exists a unique matrix $X \in \langle\Gen\rangle$ such that $\lambda = \left|\left|PXu\right|\right|^2$, then we say that $\lambda$ is \emph{unambiguous} with respect to $\qfa$ (otherwise it is ambiguous). We call $\Lambda(\qfa)$ unambiguous if every $\lambda\in \Lambda(\qfa)$ is unambiguous. 

\begin{prob}[QFA Injectivity] \label{problem:freeness} 
Given a Quantum Finite Automaton $\qfa$, is $\qfa$ injective? 
\end{prob}

\begin{prob}[QFA Scalar Ambiguity]\label{problem:ambiguity}
Given a Quantum Finite Automaton $\qfa$, is $\Lambda(\qfa)$ unambiguous?
\end{prob}

\begin{example}
Let $A = \begin{pmatrix} \frac{3}{5} & \frac{4}{5} \\ -\frac{4}{5} & \frac{3}{5} \end{pmatrix}$, $P = \begin{pmatrix} 1 & 0 \\ 0 & 0 \end{pmatrix}$ and $u = (1, 0)^T$. We thus see that $\mathcal{Q} = (P, \{A\}, u)$ is a unary $2$-state QFA. Note that $A$ represents rotations of the Euclidean plane of angle $\arccos(3/5)$, and thus we see that $\mathcal{Q}(a^k) = || P A^k u ||^2
$ is dense in $[0, 1]$ for $k \in \mathbb{N}$. Since the angle of rotation of $A$ is an irrational multiple of $\pi$, then every acceptance probability of $\mathcal{Q}$ is unique, and thus $\mathcal{Q}$ is both injective and unambiguous.
\end{example}

We show that injectivity and ambiguity are undecidable for QFA in Section~\ref{mainsec}. The reduction is from the Mixed Modification of Post's Correspondence Problem, now defined.

\begin{prob}[Mixed Modification of PCP (MMPCP)] 
Given set of letters $\Sigma = \{s_1, \ldots, s_{n}\}$, binary alphabet $\Sigma_2$, and pair of homomorphisms $h,g:\Sigma^*\to\Sigma_2^*,$ the MMPCP asks to decide whether there exists a word $w=x_1\cdots x_k\in\Sigma^+, x_i\in\Sigma$ such that
\[  
h_1(x_1)h_2(x_2)\cdots h_k(x_k)=g_1(x_1)g_2(x_2)\cdots g_k(x_k),
\]
where $h_i,g_i\in\{h,g\},$ and there exists at least one $j$ such that $h_j\neq g_j.$
\end{prob}

\begin{theorem}\hspace{-0.1cm}\cite{CKH96}\label{mmpcp9} - 
The Mixed Modification of PCP is undecidable for $|\Sigma| \geq 9$.
\end{theorem}

\begin{definition} We call an instance of the (MM)PCP a \emph{Claus instance} \footnote{Named after V.~Claus who studied this variant in \cite{Cl80}.}  if the minimal
solution words are of the form $w = s_1 x_2 x_3 \cdots x_{k-1} s_{|\Sigma|}$, where $x_2, \ldots, x_{k-1} \in \Sigma - \{s_1, s_{|\Sigma|}\}$, i.e., the minimal solution words must start with letter $s_1$, end with letter $s_{|\Sigma|}$, and all other letters are not equal to $s_1$ or $s_{|\Sigma|}$.
\end{definition}
In fact most proofs of the undecidability of (MM)PCP have this property \cite{HHH06}. Claus instances can be useful for decreasing the resources required for showing certain undecidability results, and we use this property later.

\begin{theorem}\hspace{-0.1cm}\cite{HHH06}\label{mmpcpclaus} - 
Mixed Modification PCP is undecidable for Claus instances, when $|\Sigma| \geq 9$.\footnote{The result in \cite{HHH06} states undecidability for $|\Sigma| \geq 7$ since they fix the first/last letters of a potential solution.}
\end{theorem}

\section{A mapping from arbitrary words to rational unitary matrices}\label{gammasec}

Let $\Sigma_n = \{x_1, x_2, \ldots, x_n\}$ be an $n$-letter alphabet for some $n>0$. We begin by deriving a monomorphism $\gamma: \Sigma_n^* \hookrightarrow \Q^{4 \times 4}$ such that $\gamma(w)$ is a unitary matrix for any $w \in \Sigma_n^*$. The mapping $\gamma$ will be a composition of several monomorphisms.

We first describe a monomorphism $\gamma_1$ from arbitrary sized alphabet $\Sigma_n$ to a binary alphabet $\Sigma_2$. We then show a monomorphism $\gamma_2$ from a binary alphabet $\Sigma_2$ to unit quaternions, and conclude by injectively mapping such quaternions to unitary matrices.

\begin{enumerate}
\item[$\gamma_1$:] Let $\Sigma_2 = \{a, b\}$ be a binary alphabet. We define $\gamma_1:\Sigma_n \hookrightarrow \Sigma_2^*$ by  $\gamma_1(x_k) = a^kb$ for $1 \leq k \leq n$. It is immediate that $\gamma_1$ is injective. We can trivially extend $\gamma_1$ to have domain $\Sigma_n^*$ by defining $\gamma_1(\varepsilon) = \varepsilon$ and  $\gamma_1(w_1 w_2 \cdots w_k) = \gamma_1(w_1) \gamma_1(w_2 \cdots w_n)$ recursively for a word $w_1w_2 \cdots w_k \in \Sigma^k$ and thus $\gamma_1$ is a monomorphism.

\item[$\gamma_2$:] Define mapping $\gamma_2:\Sigma_2^* \hookrightarrow \HQ(\Q)$ by $\gamma_2(a) = \left(\frac{3}{5}, \frac{4}{5}, 0, 0 \right) \cdot \mu$, $\gamma_2(b) = \left(\frac{3}{5}, 0, \frac{4}{5}, 0 \right) \cdot \mu$, and $\gamma_2(\varepsilon) = I_4$ where $I_4$ is the $4 \times 4$ identity matrix and $\gamma_2(w_1 w_2 \cdots w_k) = \gamma_2(w_1) \gamma_2(w_2 \cdots w_n)$ recursively for a word $w_1w_2 \cdots w_k \in \Sigma^k$. We may define that $\gamma_2(a^{-1}) = \gamma_2(a)^{-1}$ so that $\gamma_2(a^{-1})\gamma_2(a) = \gamma_2(a) \gamma_2(a^{-1}) = I_4$ (similarly for $\gamma_2(b^{-1})$). 

It is known that $\gamma_2$ is a monomorphism \cite{BP08quat} since such quaternions represent rotations about perpendicular axes by a rational angle (not equal to $0, \pm \frac{1}{2}, \pm 1$), thus $\gamma_2:\Sigma_2^* \hookrightarrow \HQ(\Q)$ and $\gamma_2(w_1) = \gamma_2(w_2)$ for $w_1, w_2 \in \Sigma_2^*$ implies that $w_1 = w_2$ \cite{Sw94}. 

\item[$\gamma_3$:] Define $\gamma_3:\HQ(\Q) \hookrightarrow \Q^{4 \times 4}$ by:
\begin{eqnarray}
\gamma_3((r, x, y, z)\cdot \mu) = \left(\begin{array}{cccc} r & x & y & z \\ -x & r & z & -y \\ -y & -z & r & x \\ -z & y & -x & r\end{array}\right). \label{matquatform2}
\end{eqnarray}
It is well known that $\gamma_3$ is a monomorphism in this case. Injectivity is clear, and using the rules of quaternion multiplication shows that $\gamma_3$ is a homomorphism.
\end{enumerate}

We finally define $\gamma = \gamma_3 \circ \gamma_2 \circ \gamma_1$ and thus by the above reasoning $\gamma: \Sigma_n^* \hookrightarrow \Q^{4 \times 4}$ is an injective homomorphism. Note that the matrix $\gamma(w)$ for a word $w \in \Sigma_n^*$ contains quite a lot of redundancy, and in fact can be uniquely described by just four elements (the top row) as is shown by the matrix in Eqn.~(\ref{matquatform2}). Of course, these four elements simply correspond to the four elements of the quaternion used in the construction of $\gamma$. Since these matrices are unitary (and the corresponding quaternion is of unit length), we also see that $r^2+x^2+y^2+z^2 = 1$ which will be important later.  Note also that $\gamma(w)$ is a unitary matrix since $\gamma_2$ generates a \emph{unit quaternion} (of norm $1$) in each case. 

Using $\gamma$, we can now find matrices $A, B \in \mathbb{Q}^{4 \times 4}$, such that $\gamma(w) \in \langle \{A, B\} \rangle$ for all $w \in \Sigma^*$; i.e., the value of $\gamma(w)$ lies within the semigroup generated by $\{A, B\}$. This will prove useful later since we may reason about the structure of this freely generated semigroup. 

\begin{definition}\label{freematex}
Given $\Sigma_2  = \{a, b\}$, then let:
$$
A = \gamma_3(\gamma_2(a)) = \left( \begin{array}{cccc} \frac{3}{5} & \frac{4}{5} & 0 & 0 \\ -\frac{4}{5} & \frac{3}{5} & 0 & 0 \\ 0 & 0 & \frac{3}{5} & \frac{4}{5} \\ 0 & 0 & -\frac{4}{5} & \frac{3}{5}  \end{array} \right), \,  B = 
\gamma_3(\gamma_2(b)) = \left( \begin{array}{cccc} \frac{3}{5} & 0 & \frac{4}{5} & 0 \\ 0 & \frac{3}{5} & 0 & -\frac{4}{5} \\ -\frac{4}{5} & 0 & \frac{3}{5} & 0 \\ 0 & \frac{4}{5} & 0 & \frac{3}{5}  \end{array} \right),
$$
and define $\Genn' = \langle\{A, B\}\rangle \subset \Q^{4 \times 4}$, which is a free semigroup (freely generated by $\{A, B\}$). All elements in the range of $\gamma$ thus belong to $\Genn'$. We  define $\Genn \subset \Genn'$ by $\Genn = \{\gamma(w)| w \in \Sigma_n^*\}$.
\end{definition}

\section{Freeness and ambiguity for QFA with radicals}\label{mainsec}

In order to prove that the ambiguity and freeness problems are undecidable for QFA defined over rationals (with real algebraic initial vector), we require the following (folklore) theorem. This will essentially allow us to uniquely represent a tuple of rationals as a linear sum of radicals. For completeness, we will show this simple, well-known proof of this theorem using the theory of field extensions. 

\begin{theorem} \label{fieldExtThm} 
Let $m_1, \ldots, m_n$ be pairwise coprime square-free integers. Then the set 
$\{\sqrt{m_1}, \ldots, \sqrt{m_n} \}$ is linearly independent over $\Q$. 
\end{theorem}

\begin{proof} Define $E_k=\mathbb Q(\sqrt{m_1},\ldots,\sqrt{m_k})$, where the notation stands for the field extension of $\mathbb Q$, (see \cite{Lang2002} for details)  so $E_0=\mathbb Q$ and $E_1=\mathbb Q(\sqrt{m_1})$. Clearly $[E_0:\mathbb Q]=1=2^0$ (this notation stands for the field extension degree, see \cite{Lang2002} for details), and $[E_1:\mathbb Q]=2^1$. As each element $\sqrt{m_i}$ satisfies a quadratic equation over $\mathbb Q$, the field extension degree $[E_n:\mathbb Q]$ is at most $2^n$. The theorem is proven if we can show that
$[E_n:\mathbb Q]=2^n$.

Assume the induction hypothesis true for values less than $k$. We will prove it true for $k+1$, as well, i.e., $[E_{k+1}:E_k]=2$. For this aim, we must demonstrate that $\sqrt{m_{k+1}}\notin E_k$, so let us assume the contrary, that
$$
\sqrt{m_{k+1}}\in E_k=E_{k-1}(\sqrt{m_k}),
$$
hence $\sqrt{m_{k+1}}=a+b\sqrt{m_k}$, where $a,b\in E_{k-1}$ Then
$$
m_{k+1}=a^2+m_kb^2+2ab\sqrt{m_k}.
$$
If $ab\ne0$, then $\sqrt{m_k}\in E_{k-1}$, which implies that $[E_k:E_{k-1}]=1$, a contradiction.

If $a=0$, then $\sqrt{m_{k+1}}=b\sqrt{m_k}$, and hence $\sqrt{m_k}\sqrt{m_{k+1}}=bm_k\in E_{k-1}$.
By the induction hypothesis we then have
$$
[\mathbb Q(\sqrt{m_1},\ldots,\sqrt{m_{k-1}},\sqrt{m_km_{k+1}}):\mathbb Q]=2^k,
$$
but since the last extending element belongs to $E_{k-1}$, the extension degree cannot be more than $2^{k-1}$, a contradiction. Here we actually need the assumption that the numbers are coprime, since otherwise $m_km_{k+1}$ would not necessarily be squarefree.

If $b=0$, then $\sqrt{m_{k+1}}\in\mathbb E_{k-1}$, and as above, the induction hypothesis gives
$$
[\mathbb Q(\sqrt{m_1},\ldots,\sqrt{m_{k-1}},\sqrt{m_{k+1}}):\mathbb Q]=2^k,
$$
but as the last extending element belongs to $E_{k-1}$, the extension degree cannot be more than $2^{k-1}$, a contradiction.
\end{proof}

As an illustrative example of Theorem~\ref{fieldExtThm}, given $p_1, p_2, q_1, q_2 \in \Q$, then the equality $p_1\sqrt{2} + q_1\sqrt{3} = p_2\sqrt{2} + q_2\sqrt{3}$ is true iff $p_1 = p_2$ and $q_1 = q_2$.

The following technical lemma concerns the free group $\mathcal{S}$ of quaternions generated by $\mathcal{G} = \{\gamma_2(a), \gamma_2(b)\}$ and will crucially allow us to characterise elements of $\mathcal{S}$ which differ only in the signs of one or more of their imaginary components. To formulate  this lemma we require a nonstandard inversion function defined on elements of $\mathcal{S} = \langle \mathcal{G} \rangle_{gr}$. Since $\mathcal{S}$ is free, any reduced (i.e., not containing consecutive inverses) $q_w \in \mathcal{S}$ can be uniquely written in the form 
$$
q_w = \gamma_2(a)^{k_0} \gamma_2(b)^{k_1} \gamma_2(a)^{k_2} \cdots \gamma_2(a)^{k_{n-2}} \gamma_2(b)^{k_{n-1}} \gamma_2(a)^{k_n},
$$
where $k_0, k_n \in \mathbb{Z}$ and $k_1, \ldots, k_{n-1} \in \mathbb{Z} - \{0\}$, i.e., an alternating product of either positive or negative powers of $\gamma_2(a)$ and $\gamma_2(b)$ which may start and end with either element. We define the following three functions:
\begin{enumerate}
\item[i)] $\lambda_a(q_w) = \gamma_2(a)^{-k_0} \gamma_2(b)^{k_1} \gamma_2(a)^{-k_2} \cdots \gamma_2(a)^{-k_{n-2}} \gamma_2(b)^{k_{n-1}} \gamma_2(a)^{-k_n}$;
\item[ii)] $\lambda_b(q_w) = \gamma_2(a)^{k_0} \gamma_2(b)^{-k_1} \gamma_2(a)^{k_2} \cdots \gamma_2(a)^{k_{n-2}} \gamma_2(b)^{-k_{n-1}} \gamma_2(a)^{k_n}$;
\item[iii)]$\lambda_{a,b}(q_w) = \gamma_2(a)^{-k_0} \gamma_2(b)^{-k_1} \gamma_2(a)^{-k_2} \cdots \gamma_2(a)^{-k_{n-2}} \gamma_2(b)^{-k_{n-1}} \gamma_2(a)^{-k_n}$.
\end{enumerate}
These three functions thus invert all $\gamma_2(a)$ elements in a product for $\lambda_a$, all  $\gamma_2(b)$ elements in a product for $\lambda_b$ and both  $\gamma_2(a)$ and $\gamma_2(b)$ elements in a product for $\lambda_{a,b}$. As an example, if $q_w = \gamma_2(a)^3\gamma_2(b)^2\gamma_2(a)^{-4}\gamma_2(b)$, then $\lambda_a(q_w) = \gamma_2(a)^{-3}\gamma_2(b)^2\gamma_2(a)^{4}\gamma_2(b)$, $\lambda_b(q_w) = \gamma_2(a)^{3}\gamma_2(b)^{-2}\gamma_2(a)^{-4}\gamma_2(b)^{-1}$ and $\lambda_{a, b}(q_w) = \gamma_2(a)^{-3}\gamma_2(b)^{-2}\gamma_2(a)^{4}\gamma_2(b)^{-1}$. Bizarre as such a definition may appear, it allows us to exactly characterize those elements of $\mathcal{S}$ which differ only in the sign of one or more of their imaginary components, as we now show.

\begin{lemma}\label{fourthneg}
Given a quaternion $q_w = \gamma_2(w) = (r, x, y, z) \cdot \mu \in \langle \gamma_2(a), \gamma_2(b)\rangle_{gr}$ with $w = w_1 w_2 \cdots w_{|w|}$, each $w_i \in (\Sigma_2 \cup \Sigma_2^{-1})$ and $\Sigma_2 = \{a, b\}$, then: 
\begin{enumerate}
\item[i)] $q_{w^R} = \gamma_2(w^R) =  (r,x,y,-z) \cdot \mu$;
\item[ii)] $\lambda_{a}(q_{w}) = (r,-x,y,-z) \cdot \mu$;
\item[iii)] $\lambda_{b}(q_{w}) = (r,x,-y,-z) \cdot \mu$;
\item[iv)] $\lambda_{a, b}(q_{w}) = (r,-x,-y,z) \cdot \mu$.
\end{enumerate}
\end{lemma}
\begin{proof}
We proceed via induction. For the base case, when $w=\varepsilon$, then $q_w = (1, 0, 0, 0) \cdot \mu$ and  $q_{w^{R}} =\lambda_{a}(q_{w})= \lambda_{b}(q_{w})= \lambda_{a,b}(q_{w})=  (1, 0, 0, 0) \cdot \mu$ and so the properties (trivially) hold. For the induction hypothesis, assume $i)$ -- $iv)$ are true for $q_w$. We handle each property individually. 

\noindent {\bf i)} By assumption, $q_{w^R} = (r, x, y, -z)\cdot\mu$. Since $\gamma_2(a) = \left(\frac{3}{5}, \frac{4}{5}, 0, 0 \right) \cdot \mu$ and $\gamma_2(b) = \left(\frac{3}{5}, 0, \frac{4}{5}, 0 \right) \cdot \mu$, by the rules of quaternion multiplication, we see that:
\begin{eqnarray*}
\gamma_2(a)\cdot q_w & = & \frac{1}{5}\left(3r-4x, 3x+4r, 3y-4z, 3z+4y\right)\cdot\mu, \\
q_{w^R} \cdot \gamma_2(a) & = & \frac{1}{5}\left(3r-4x, 3x+4r, 3y-4z, -3z-4y\right)\cdot\mu
\end{eqnarray*}
Note that the fourth component is negated as expected. In a similar way, we also see that:
\begin{eqnarray*}
\gamma_2(b)\cdot q_w & = & \frac{1}{5}\left(3r-4y,3x+4z, 3y+4r, 3z-4x\right)\cdot\mu, \\
q_{w^R} \cdot\gamma_2(b) & = & \frac{1}{5}\left(3r-4y,3x+4z, 3y+4r, -3z+4x\right)\cdot\mu
\end{eqnarray*}
with negated fourth element. Since $\gamma_2(a^{-1}) = \left(\frac{3}{5}, -\frac{4}{5}, 0, 0 \right) \cdot \mu$ and $\gamma_2(b^{-1}) = \left(\frac{3}{5}, 0, -\frac{4}{5}, 0 \right) \cdot \mu$, then the property of the fourth element being negated is also clearly true for $\gamma_2(c^{-1}) \cdot q_w$ and $q_{w^R} \cdot \gamma_2(c^{-1})$ for $c \in \{a, b\}$. The other properties are similar, we give a brief proof of each.

\noindent {\bf ii)} By the induction hypothesis, $\lambda_{a}(q_{w}) = (r,-x,y,-z) \cdot \mu$ and thus:
\begin{eqnarray*}
q_w \cdot\gamma_2(a) & = & \frac{1}{5}\left(3r - 4x, 3x+ 4r, 3y + 4z, 3z - 4y\right)\cdot\mu, \\
\lambda_{a}(q_{w}) \cdot \gamma_2(a)^{-1} & = & \frac{1}{5}\left(3r - 4x, -3x - 4r, 3y + 4z, -3z+4y\right)\cdot\mu,
\end{eqnarray*}
with the second and fourth components negated as required. Also,
\begin{eqnarray*}
q_w \cdot\gamma_2(a)^{-1} & = & \frac{1}{5}\left(3r + 4x, 3x - 4r, 3y - 4z, 3z+ 4y\right)\cdot\mu, \\
\lambda_{a}(q_{w}) \cdot \gamma_2(a) & = & \frac{1}{5}\left(3r + 4x, -3x+4r, 3y - 4z, -3z- 4y\right)\cdot\mu,
\end{eqnarray*}
as expected. Right multiplication of $q_w$ and $\lambda_{a}(q_{w})$ by either $\gamma_2(b)$ or $\gamma_2(b)^{-1}$ retains the given structure, as is not difficult to calculate.

\noindent {\bf iii)} By the induction hypothesis, $\lambda_{b}(q_{w}) = (r,x,-y,-z) \cdot \mu$ and thus:
\begin{eqnarray*}
q_w \cdot\gamma_2(b) & = & \frac{1}{5}\left(3r - 4y, 3x - 4z, 3y+ 4r , 3z+ 4x\right)\cdot\mu, \\
\lambda_{b}(q_{w}) \cdot \gamma_2(b)^{-1} & = & \frac{1}{5}\left(3r - 4y, 3x - 4z, -3y - 4r, -3z - 4x\right)\cdot\mu,
\end{eqnarray*}
with the third and fourth components negated as required. Also,
\begin{eqnarray*}
q_w \cdot\gamma_2(b) ^{-1} & = & \frac{1}{5}\left(3r + 4y, 3x + 4z, 3y - 4r, 3z - 4x\right)\cdot\mu, \\
\lambda_{b}(q_{w}) \cdot \gamma_2(b) & = & \frac{1}{5}\left(3r + 4y, 3x + 4z, -3y+ 4r , -3z+ 4x\right)\cdot\mu,
\end{eqnarray*}
as expected. Right multiplication of $q_w$ and $\lambda_{b}(q_{w})$ by either $\gamma_2(a)$ or $\gamma_2(a)^{-1}$ retains the given structure, as is not difficult to calculate.

\noindent {\bf iv)} By the induction hypothesis, $\lambda_{a, b}(q_{w}) = (r,-x,-y,z) \cdot \mu$ and thus:
\begin{eqnarray*}
\lambda_{a, b}(q_{w}) \cdot \gamma_2(a) & = & \frac{1}{5}\left(3r + 4x, -3x+4r, -3y+4z, 3z+4y\right)\cdot\mu, \\
\lambda_{a, b}(q_{w}) \cdot \gamma_2(b) & = & \frac{1}{5}\left(3r + 4y, - 3x - 4z, -3y+4r, 3z - 4x\right)\cdot\mu, \\
\lambda_{a, b}(q_{w}) \cdot \gamma_2(a)^{-1} & = & \frac{1}{5}\left(3r - 4x, -3x - 4r, - 3y - 4z, 3z - 4y\right)\cdot\mu, \\
\lambda_{a, b}(q_{w}) \cdot \gamma_2(b)^{-1} & = & \frac{1}{5}\left(3r - 4y, -3x+4z, -3y - 4r, 3z+ 4x\right)\cdot\mu,
\end{eqnarray*}
with the second and third components of each product negated with relation to $q_w \cdot\gamma_2(a)^{-1}$, $q_w \cdot\gamma_2(b)^{-1}$, $q_w \cdot\gamma_2(a)$ and $q_w \cdot\gamma_2(b)$ (resp.) as required.
\end{proof}

The following lemma allows us to represent a quaternion (and its corresponding rotation matrix) by using only absolute values and will be crucial later.

\begin{lemma}\label{uniquequat2}
Given a word $w \in \Sigma_k^*$, then $\gamma_2(\gamma_1(w)) = (r, x, y, z)\cdot \mu$ is uniquely determined by $(|r|, |x|, |y|)$. All matrices $\gamma(w) \in \Gamma$ are similarly uniquely determined by $$(|\gamma(w)_{1,1}|, |\gamma(w)_{1,2}|, |\gamma(w)_{1,3}|),$$
i.e., by the absolute values of the first three elements of the top row of the matrix.
\end{lemma}

\begin{remark} An embedding
\[
a\mapsto \frac{1}{5}\left(\begin{array}{cc} 3 & -4\\ 4 & 3 \end{array}\right),\quad
b\mapsto \frac{1}{5}\left(\begin{array}{cc} 3 &  4i\\ 4i & 3 \end{array}\right)
\] 
of the binary alphabet using unitary  $2\times 2$-matrices was introduced in \cite{Hir}, but unfortunately this embedding does not enjoy the unique first row property of Lemma \ref{uniquequat2}, as trivially seen from the images of $a$ and $b$. 
\end{remark}

\begin{proof}[Proof of Lemma \ref{uniquequat2}]
Another way to state this lemma is that if we have $u = u_1u_2 \cdots u_t$ and $v = v_1v_2 \cdots v_{t'}$ with each $u_i, v_i \in \Sigma_k^*$, such that  $\gamma_2(\gamma_1(u)) = (a_1, b_1, c_1, d_1)\cdot \mu$, $\gamma_2(\gamma_1(v)) = (a_2, b_2, c_2, d_2)\cdot \mu$ and $(|a_1|, |b_1|, |c_1|) = (|a_2|, |b_2|, |c_2|)$, then $t = t'$ and $u_i = v_i$ for all $1 \leq i \leq t$. A similar property holds for the top row of the unitary matrices when applying $\gamma_3$ to these elements\footnote{We highlight here the importance of $\gamma_1$ since $\{\gamma_2(a), \gamma_2(b)\}$ does not have the property of being uniquely determined by the absolute value of its elements as $\gamma_2(a)\gamma_2(b) = \frac{1}{25} (9, 12, 12, -16)\mu$ and $\gamma_2(b)\gamma_2(a) = \frac{1}{25}(9, 12, 12, 16)\mu$ illustrate. We see that $ba$ is not in the range of $\gamma_1$ however.}. We first prove that each matrix is uniquely determined by $(|r|, |x|, |y|, |z|)$ and then note that since $r^2+x^2+y^2+z^2 = 1$, then in fact $(|r|, |x|, |y|)$ is sufficient to uniquely define the matrix. We shall now prove these results.

By definition, $\gamma_2:\Sigma_2^* \hookrightarrow \HQ(\Q)$ maps to a free submonoid  $\mathcal{S}$ of $\HQ(\Q)$ generated by $\mathcal{G} = \{\gamma_2(a), \gamma_2(b)\}$ with $\gamma_2(a) = \left(\frac{3}{5}, \frac{4}{5}, 0, 0 \right) \cdot \mu$ and $\gamma_2(b) = \left(\frac{3}{5}, 0, \frac{4}{5}, 0 \right) \cdot \mu$. As shown in Section~\ref{gammasec}, $\gamma_2 \circ \gamma_1 : \Sigma_n^* \hookrightarrow \HQ(\Q)$; i.e., $\gamma_2 \circ \gamma_1 $ is an injective homomorphism. Let $\Gamma' = \{\gamma_2(\gamma_1(w')) | w' \in \Sigma_n^*\} \subseteq \HQ(\Q)$. Clearly then, $\Gamma'$ is freely generated by $\{\gamma_2(\gamma_1(w')) | w' \in \Sigma_n\}$ by the injectivity of $\gamma_2 \circ \gamma_1$.

Let $q_w =  \gamma_2(\gamma_1(w)) = (r, x, y, z) \cdot \mu \in \Gamma' \subseteq \mathcal{S}$ and define $Q_w = \{(\pm r, \pm x, \pm y, \pm z) \cdot \mu\}$, thus $|Q_w| = 16$. We will now show that $q' \not\in \Gamma'$ for all $q' \in Q_w - \{q_w\}$ which proves the lemma.

Since (unit) quaternion inversion simply involves negating all imaginary components, then using the identities of Lemma~\ref{fourthneg}, we can derive that $q_w^{-1} = (r, -x, -y, -z)\mu$, $\lambda_{a}(q_w)^{-1} =  (r, x, -y, z)\mu$ and $\lambda_{b}(q_w)^{-1} =  (r, -x, y, z)\mu$ which we summarize in the following table.

{\renewcommand{\arraystretch}{1.2}
\begin{tabular}{ |c|l|c|l|}
\hline
$q_w$ & $(r, x, y, z)\mu$ & $q_{w}^{-1}$ & $(r, -x, -y, -z)\mu$  \\
\hline
$\lambda_a(q_w)$ & $(r, -x, y, -z)\mu$ & $\lambda_{a}(q_w)^{-1}$ & $(r, x, -y, z)\mu$ \\
\hline
$\lambda_b(q_w)$ & $(r, x, -y, -z)\mu$ & $\lambda_{b}(q_w)^{-1}$ & $(r, -x, y, z)\mu$ \\
\hline
$\lambda_{a,b}(q_w)$ & $(r, -x, -y, z)\mu$ & $q_{w^R}$ & $(r, x, y, -z)\mu$ \\
\hline
\end{tabular}
}

We might also notice other identities, such as $q_{w^R} = \lambda_{a, b}(q_w)^{-1}$ which is clear from the definition of $\lambda_{a, b}$. Note that this table covers $8$ elements of $Q_w$.

Note that $q_w$ belongs (by definition) to $\Gamma' = (\gamma_2(a)^+\gamma_2(b))^+ = \{\gamma_2(\gamma_1(w')) | w' \in \Sigma_n^*\} \subseteq \mathcal{S}$. Since $\langle\gamma_2(a), \gamma_2(b) \rangle_{gr}$ is a free group, this means that no reduced element of $\mathcal{S}$ is equal to a product with a nontrivial\footnote{Recall that reduced meaning the element contains no consecutive inverse elements and nontrivial meaning we ignoring any such element adjacent to its multiplicative inverse.} factor $\gamma_2(a)^{-1}$ or $\gamma_2(b)^{-1}$. Each element in the above table contains at least one irreducible factor $\gamma_2(a)^{-1}$ or $\gamma_2(b)^{-1}$, excluding $q_w$ and $q_{w^R}$. Note however that $q_{w^R}$ trivially does not belong to $\Gamma' = (\gamma_2(a)^+\gamma_2(b))^+$ since it necessarily begins with irreducible factor $\gamma_2(b)$.

Finally, to cover the remaining $8$ elements of $Q_w$, we consider the \emph{free group} $\mathcal{S}_{\textup{gr}} = \langle \{\gamma_2(a), \gamma_2(b)\} \rangle_{gr}$. For any $q'_w \in \mathcal{S}_{\textup{gr}}$ then $-q'_w \not\in \mathcal{S}_{\textup{gr}}$ since $\mathcal{S}_{\textup{gr}}$ is free. This holds since if $-q'_w \in \mathcal{S}$, then $-1 \in \mathcal{S}$ (because $(q'_w)^{-1} \in \mathcal{S}$), which gives a nontrivial identity $-1^2 = 1$ in $\mathcal{S}_{\textup{gr}}$ (a contradiction). 

This covers all sixteen possible elements of $Q_w$ and shows that $q_w$ is the only member of $Q_w$ belonging to $\Gamma'$. By the definition of $\gamma_3:\HQ(\Q) \hookrightarrow \Q^{4 \times 4}$, then also all matrices $\gamma(w) \in \Gamma$ are uniquely determined by their top row $(|\gamma(w)_{1,1}|, |\gamma(w)_{1,2}|, |\gamma(w)_{1,3}|, |\gamma(w)_{1,4}|)$
as required.

Finally then, as mentioned previously, we note that $r^2+x^2+y^2+z^2 = 1$, and thus $(|r|, |x|, |y|)$ is sufficient to uniquely define each matrix, proving the result.
\end{proof}

\begin{theorem}\label{basistheorem2}
The injectivity problem for measure-once quantum finite automata is undecidable for $8$ states over an alphabet of size $17$.
\end{theorem}

\begin{proof}
We will encode an instance $(h, g)$ of the mixed modification of Post's Correspondence Problem into a QFA $\qfa$ so that there exists a solution to the instance if and only if $\qfa$ is not injective (i.e. $\qfa(w_1) = \qfa(w_2)$ for some $w_1 \neq w_2$).

Let $\Sigma=\{x_1,x_2,\dots,x_{n-2}\}$ and $\Delta=\{x_{n-1},x_n\}$ be distinct alphabets and $h, g:\Sigma^* \to \Delta^*$ be an instance of the mixed modification of PCP and let $\Sigma_n = \Sigma \cup \Delta$. The naming convention will become apparent below, but intuitively we will be applying $\gamma$, from Section~\ref{gammasec} to both the input and output alphabets.

Recall that we showed the injectivity of $\gamma$ in Section~\ref{gammasec}, and thus have a monomorphism $\gamma: \Sigma_n^* \hookrightarrow \Q^{4 \times 4}$. We define a function $\varphi: \Sigma_n^* \times \Sigma_n^* \hookrightarrow \Q^{8 \times 8}$ by 
$$\varphi(w_1, w_2) = \gamma(w_1) \oplus \gamma(w_2).$$

We may note that  $\varphi(w_1, w_2)$ remains a unitary matrix since $\gamma(w_i)$ is unitary and the direct sum of unitary matrices is unitary as in Lemma~\ref{kronprop}. We now define $\Gen = \{\varphi(x_i, h(x_i)), \varphi(x_i, g(x_i)) | x_i \in \Sigma\} \subset \Q^{8 \times 8}$.

Let $p_i$ denote the i'th prime number and define $u = \frac{u_1 \oplus u_2}{\sqrt{\sum_{i=1}^{6}\sqrt{p_i}}} \in \A^{8}$, where $u_1 = (\sqrt[4]{p_1}, \sqrt[4]{p_2}, \sqrt[4]{p_3}, 0)^T$ and $u_2 = (\sqrt[4]{p_4}, \sqrt[4]{p_5}, \sqrt[4]{p_6}, 0)^T$, noting that $u$ is thus a unit length vector. Note that $\A^8$ denotes a $8$-tuple of elements from $\A$ (real algebraic numbers).

Let $P = P_1 \oplus P_1  \in \Q^{8 \times 8}$ where $P_1 = 1 \oplus {\bf 0}_3$ and ${\bf 0}_3$ is the $3 \times 3$ zero matrix; thus $P$ has a $1$ in positions $P_{1,1}$ and $P_{5,5}$ and zero elsewhere. Note that $P^2 = P$ and $P$ is a projection matrix.

The QFA $\mathcal{Q}$ is thus defined by the triple $\mathcal{Q} = (P, \Gen, u)$ and we now prove the claim of the theorem.

Let $X = X_{i_1} \cdots X_{i_p} = \varphi(x_{i_1}, f_{i_1}(x_{i_1})) \cdots \varphi(x_{i_p}, f_{i_p}(x_{i_p}))$, with $f_{i_k} \in \{g, h\}$ for $1 \leq k \leq p$ be one factorization of a matrix $X \in \Gen$. Define $x = x_{i_1} \cdots x_{i_p}$ and $f(x) = f_{i_1}(x_{i_1}) \cdots f_{i_p}(x_{i_p})$. Then we see that:

\begin{eqnarray}
\left|\left| PXu\right|\right|^2 & =  & \left|\left| \frac{(P_1 \oplus P_1)(\gamma(x) \oplus \gamma(f(x)))(u_1 \oplus u_2)}{\sqrt{\sum_{i=1}^{6}\sqrt{p_i}}} \right|\right|^2 \label{eqq0} \\
& = & \left|\left| \frac{P_1 \gamma(x) u_1 \oplus P_1 \gamma(f(x)) u_2}{\sqrt{\sum_{i=1}^{6}\sqrt{p_i}}} \right|\right|^2 \label{eqq1} \\
& =  & \left( \sqrt{\frac{\sum_{j=1}^3(\gamma(x)_{1,j} \sqrt[4]{p_j})^2 + \sum_{j=1}^3 (\gamma(f(x))_{1,j} \sqrt[4]{p_{3+j}})^2}{\sqrt{\sum_{i=1}^{6}\sqrt{p_i}}} }\right)^2 \label{eqq3} \\
& =  & \frac{\sum_{j=1}^3\gamma(x)_{1,j}^2 \sqrt{p_j} + \sum_{j=1}^3 \gamma(f(x))_{1,j}^2 \sqrt{p_{3+j}}}{\sqrt{\sum_{i=1}^{6}\sqrt{p_i}}}. \label{eqq4b}
\end{eqnarray}
In the above, we used the fact that $P_1\gamma(x)e_j = \gamma(x)_{1,j}$ as well as the properties of Kronecker products from Lemma~\ref{kronprop}. Assume that matrix $X$ has two distinct factorizations $X = X_{i_1} \cdots X_{i_p} = X_{j_1} \cdots X_{j_q} \in \Gen^+$ and $p \neq q$ or $X_{i_k} \neq X_{j_k}$ for some $1 \leq k \leq p$, such that
$$\left|\left| PXu\right|\right|^2 = \left|\left| PX_{i_1} \cdots X_{i_p}u\right|\right|^2 = \left|\left| PX_{j_1} \cdots X_{j_q}u\right|\right|^2,
$$
and thus $\mathcal{Q}$  is not injective since $\qfa(w_{i_1}\cdots w_{i_p}) = \qfa(w_{j_1}\cdots w_{j_q})$ with $w_{i_1}\cdots w_{i_p} \neq w_{j_1}\cdots w_{j_q}$, where each $w_\ell \in \Sigma$ and $w_\ell \mapsto X_{\ell}$.  Assume then that $X = X_{j_1} \cdots X_{j_q} = \varphi(x_{j_1}, f'_{j_1}(x_{j_1})) \cdots \varphi(x_{j_q}, f'_{j_q}(x_{j_q}))$, with $f'_{j_k} \in \{g, h\}$ for $1 \leq k \leq q$ is another factorization of $X$ and define $x' = x_{j_1} \cdots x_{j_q}$ and $f'(x') = f'_{j_1}(x_{j_1}) \cdots f'_{j_q}(x_{j_p})$ with each $f'_{j_k} \in \{g,h\}$. Note that in Eqn.~(\ref{eqq4b}) the denominator is constant and thus when determining equality  $\left|\left| PX_{i_1} \cdots X_{i_p}u\right|\right|^2 = \left|\left| P X_{j_1} \cdots X_{j_q}u\right|\right|^2$ we may ignore it. By Lemma~\ref{uniquequat2}, each $\gamma(w)$ is uniquely determined by the tuple $(|\gamma(w)_{1, 1}|, |\gamma(w)_{1, 2}|, |\gamma(w)_{1, 3}|)$.  Since each $p_j$ is squarefree, for $1 \leq j \leq 6$, then by Theorem~\ref{fieldExtThm}, the following equation is satisfied if and only if $\gamma(x) = \gamma(x')$ and $\gamma(f(x)) = \gamma(f'(x'))$:
\begin{eqnarray*}
& & \sum_{j=1}^3 \gamma(x)_{1,j}^2 \sqrt{p_j} + \sum_{j=1}^3 \gamma(f(x))_{1,j}^2 \sqrt{p_{3+j}}  \\
& = & \sum_{j=1}^3 \gamma(x')_{1,j}^2 \sqrt{p_j} + \sum_{j=1}^3 \gamma(f'(x'))_{1,j}^2 \sqrt{p_{3+j}}.
\end{eqnarray*}

Finally, note that $\gamma(x) = \gamma(x')$ if and only if $x = x'$ since $\gamma$ is an injective homomorphism. As before, let $x = x_{i_1} \cdots x_{i_p}$, then  $\gamma(f(x)) =  f_{i_1}(x_{i_1}) \cdots f_{i_p}(x_{i_p}) =  f_{j_1}(x_{i_1}) \cdots f_{j_p}(x_{i_p}) = \gamma(f'(x'))$ with some $f_{i_k} \neq f_{j_k}$ for $1 \leq k \leq p$ if and only if the instance of the MMPCP has a solution and the first part of the proof is done.  

If the MMPCP is undecidable for Claus instances with an alphabet of size $n'$ (see Theorem~\ref{mmpcpclaus}), then the undecidability of the current theorem holds for $|\mathcal{G}| \geq 2n'$. We now prove that the result holds for $|\mathcal{G}| \geq 2n'-1$. Let $\Sigma = \{x_1, \ldots, x_{n'}\}$. Since $h, g$ is a Claus instance, any solution word $w$ is of the form $w = x_1 w' x_{n'}$, with $w' \in (\Sigma - \{x_1, x_{n'}\})^*$. By symmetry, we may assume that $h_1 = h$ and by the proof in \cite{HHH06}, $g_i = g$ and $h_i = h$ for all $1 \leq i \leq n'$.  Clearly then, one of $h(x_{n'})$ and $g(x_{n'})$ is a proper suffix of the other (assume that $g(x_{n'})$ is a suffix of $h(x_{n'})$; the opposite case is similar). Now, redefine $u' = \gamma(x_{n'}, g(x_{n'}))u$, remove the matrix corresponding to  $g(x_{n'})$ from $\mathcal{G}$ and redefine the matrix corresponding to $h(x_{n'})$ by $h'(x_{n'}) = \gamma(x_{n'}, h(x_{n'})g(x_{n'})^{-1})$. 
Since $g(x_{n'})$ is a proper suffix of $h(x_{n'})$, then $h(x_{n'})g(x_{n'})^{-1}$  is the prefix of $h(x_{n'})$ after removing the common suffix with $g(x_{n'})$. This means that an ambiguous scalar only exists if there exists a solution to the instance of MMPCP and we had reduced the alphabet size by $1$. MMPCP is undecidable for instances of size $9$ (Theorem~\ref{mmpcpclaus}), thus the undecidability holds for MO-QFA with $8$ states and an alphabet size of $17$.
\end{proof}

\begin{corollary}\label{basistheorem3}
The ambiguity problem for measure-once quantum finite automata is undecidable for $9$ states over an alphabet of size $17$.
\end{corollary}
\begin{proof}
The corollary follows from the proof of Theorem~\ref{basistheorem2}. We notice that if there exists a solution to the encoded instance of the MMPCP, then some matrix $U$ has two distinct factorizations over $\Gen$ and therefore there exist two distinct matrix products giving the same scalar. Our technique in this corollary is to make these two factorizations produce distinct matrices $U_1$ and $U_2$, such that they still lead to the same scalar. This is simple to accomplish by redefining the projection matrix $P$ as $P' = P \oplus 0$, redefining the initial vector $u$ as $u' = u \oplus 0$ and for each matrix $M \in \Gen - \{\varphi(x_1, h(x_1))\}$, we extend  $M$ as $M' = M \oplus 1$ and let $\varphi(x_1, h(x_1))$ be redefined as $\varphi(x_1, h(x_1)) \oplus -1$. In this case, any matrix product containing $X_{1} \oplus -1$ will have $-1$ in the bottom right element, otherwise the bottom right element is $1$. Since we encode a Claus instance of MMPCP, one factorization has $-1$ in this case, and the other has $1$, and thus we always have distinct matrices. If no solution exists, then each matrix leads to a unique scalar anyway. 

Note that we increased the number of states of the MO-QFA by $1$ and also note that the acceptance probability is unaffected by the above modifications since the projection matrix was increased by a zero row/column.
\end{proof}

\begin{remark} In a final stage of the proof of Theorem \ref{basistheorem2}, we introduced numbers
$\sqrt{p_i}$, where $p_i$ stand for the $i$'th prime number, to ensure the injectivity of mapping
$X_w\mapsto\norm{PX_w u}^2$. It is worth noticing that the irrational numbers occur only in the initial vector, but  it is however justified to criticize this solution and ask whether the undecidability of injectivity is possible only with irrational numbers. In the next section we discuss the difficulties of removing this restriction and then provide a solution.
\end{remark}

\section{Injectivity without radicals}\label{nonrad}

In the previous section, we showed how to construct a QFA with matrices $X_{w}$, and also how the  conclusion  depended on the injectivity of mapping $X_w\mapsto\norm{PX_w u}^2$.
For an alternative injectivity construction, we will demonstrate how to build a QFA computing a multivariate polynomial on the matrix elements. We begin by deriving an injective polynomial whose  domain is tuples of $n$ rational numbers of a specific form,  before using the derived polynomial to show the undecidability of the injectivity problem for QFA defined entirely over rationals, although with significantly more states than in Theorem~\ref{basistheorem2} which used algebraic numbers for the initial vector.

\subsection{Injective polynomials over tuples of rationals}

In order to avoid the use of radicals, we could try to construct an injective multivariate polynomial 
on the rational numbers. However, in this route, we face already a very hard question: does there exist a multivariate polynomial $f\in\mathbb N[x_1,\ldots,x_6]$ which is injective on rational numbers? 

This is of course a special case of a more general question for an arbitrary $k$: does there exist a multivariate polynomial $f\in\mathbb N[x_1,\ldots,x_k]$ so that
$f:\mathbb Q^k\to\mathbb Q$ is an injection? However, it is easy to see that if we have an injection in the special case $k=2$, then we can extend it to the general case recursively:
\[
f_3(x,y,z)=f(x,f(y,z)),\quad f_4(x,y,z,w)=f(x,f(y,f(z,w))),
\]
and so on. For the purposes of this article, it is also clear that in order to remove the radicals, we can use a polynomial injection only on the positive rationals, if such exists. Hence we can ask: does there exist a multivariate polynomial $f\in\mathbb N_0[x_1,\ldots,x_k]$ so that
$f:\mathbb Q_{\ge 0}^k\to\mathbb Q_{\ge 0}$ is an injection? As mentioned above, we can restrict to the case $k=2$ here also.

The existence of such a polynomial injection is not at all a straightforward issue. We may recall that the famous {\em Cantor pairing} defined as
\[
C:\mathbb N_0\times\mathbb N_0\to\mathbb N_0,
C(x,y)=\frac12(x+y+1)(x+y)+x
\]
is a bijection, $C(0,0)=0$, $C(0,1)=1$, $C(1,0)=2$, $C(0,2)=3$, $C(1,1)=4$ etc.

The Cantor pairing is unique in the sense that it is almost straightforward to discover, but it has been very long known that no degree $2$ polynomial bijections $\mathbb N_0\times\mathbb N_0\to\mathbb N_0$ essentially different from $C(x,y)$ and $C(y,x)$ exist \cite{FueterPolya}, \cite{Vsemirnov}, and more recently that
no  degree $>2$ polynomial bijection $\mathbb N_0\times\mathbb N_0\to\mathbb N_0$ exists \cite{Adriaans}.

Unfortunately, the Cantor Pairing is not our desired injection $\mathbb Q\times\mathbb Q\to\mathbb Q$, as the counterexample
\[
C\left(\frac{2}{25},\frac{11}{25}\right)=\frac{297}{625}=C\left(\frac{3}{25},\frac{9}{25}\right).
\]
shows.

We present here our own injectivity proof on a very narrow set of rational numbers.

\begin{theorem} Let $\Lambda=\{\frac{a}{5^k}\mid a,k\in\mathbb N_0,a< 5^k\}$. Then
\[
f:\Lambda\times\Lambda\to 25\Lambda
\]
defined by $f(x,y)=(x^4+y^4)^3+x^4$ is an injection.
\end{theorem}

\begin{remark} Note that by definition, the set $\Lambda$ consists of positive rational numbers having only powers of $5$ in the denominator.
We can estimate the value of $f(x,y)$ simply as,
\[
\abs{(x^4+y^4)^3+x^4}\le (1+1)^3+1=9<25,
\]
so the image is certainly in the set $25\Lambda$. An injection $\Lambda\times\Lambda\to\Lambda$ can hence be obtained simply by introducing a renormalization factor $\frac{1}{25}$ to $f$. For simplicity, we however present the injectivity proof in the form stated in the theorem. 
\end{remark}

Before presenting the proof, we would like to remind the reader of the following notations from elementary number theory:
\begin{definition} For any positive integer $n$, notation $\mathbb Z_n$ stands for the residue class ring $\mathbb Z_n=\mathbb Z/n\mathbb Z$. Notation $\mathbb Z_n^*\subset\mathbb Z_n$ means the multiplicative group in $\mathbb Z_n$ consisting exactly of those elements having a multiplicative inverse. It is a well-known fact that $\mathbb Z_n^*=\{a+n\mathbb Z\mid \gcd(a,n)=1\}$ and that the cardinality of $\mathbb Z_n^*$ is given by the {\em Euler totient function}
$\varphi(n)=n(1-\frac{1}{p_{i_1}})\ldots(1-\frac{1}{p_{i_n}})$, where
$n=p_{i_1}^{k_1}\ldots p_{i_n}^{k_n}$ is the prime factorization of $n$.
\end{definition}

\begin{proof}  
Assume that
\begin{equation}\label{injective01}
f\left(\frac{a}{5^k},\frac{b}{5^k}\right)=f\left(\frac{c}{5^k},\frac{d}{5^k}\right),
\end{equation}
where $a$, $b$, $c$, $d$, and $k\in\mathbb N_0$. Without loss of generality we can assume that at least one of the integers $a$, $b$, $c$, and $d$ is not divisible by $5$, otherwise we could reduce by $5$ to get a similar presentation with a smaller value of $k$.

For brevity, denote $\alpha=(\frac{a}{5^k})^4$, $\beta=(\frac{b}{5^k})^4$, $\gamma=(\frac{c}{5^k})^4$, and $\delta=(\frac{d}{5^k})^4$, $S_1=\alpha+\beta$, and
$S_2=\gamma+\delta$. Using these notations (\ref{injective01}) becomes
\begin{equation}\label{injective02}
S_1^3+\alpha=S_2^3+\gamma.
\end{equation}
If now $S_1=S_2$, then clearly $\alpha=\gamma$, which directly implies that also $\beta=\delta$.
As each $a$, $b$, $c$, and $d>0$, this would also imply that $a=c$ and $b=d$.

We can therefore continue with the assumption that $S_1\ne S_2$, and without loss of generality also that
\[
\alpha+\beta=S_1>S_2=\gamma+\delta.
\]
Assume first that $S_1$ and $S_2$ are at least a unit away from each other, meaning that $\Delta=S_1-S_2\ge 1$. Equation (\ref{injective02}) can then be rewritten and estimated as
\begin{eqnarray*}
& &(S_2+\Delta)^3+\alpha=S_2^3+\gamma\\
&\Leftrightarrow& S_2^3+3S_2^2\Delta+3S_2\Delta^2+\Delta^3+\alpha=S_2^3+\gamma\\
&\Leftrightarrow&\gamma-\alpha=3S_2^2\Delta+3S_2\Delta^2+\Delta^3\\
&\ge&3S_2^2+3S_2+1=2S_2^2+S_2^2+2S_2+1+S_2\\
&=&2S_2^2+(S_2+1)^2+S_2> S_2=\gamma+\delta
\end{eqnarray*}
Thus we get an inequality $\gamma-\alpha>\gamma+\delta$, which is a contradiction, since $\alpha$, $\delta\ge 0$.

We will then consider the case $\abs{S_1-S_2}<1$, which can be rewritten as
\begin{eqnarray}
& &
\abs{\big(\frac{a}{5^k}\big)^4+\big(\frac{b}{5^k}\big)^4-\Big(\big(\frac{c}{5^k}\big)^4+\big(\frac{d}{5^k}\big)^4\Big)}<1\notag \\
&\Rightarrow&
\abs{a^4+b^4-(c^4+d^4)}<5^{4k}<5^{8k}.\label{injective03}
\end{eqnarray}
If now (\ref{injective01}) holds, then
\begin{eqnarray}
& &\Big(\big(\frac{a}{5^k}\big)^4+\big(\frac{b}{5^k}\big)^4\Big)^3+
\big(\frac{a}{5^k}\big)^4
=\Big(\big(\frac{c}{5^k}\big)^4+\big(\frac{d}{5^k}\big)^4\Big)^3+
\big(\frac{c}{5^k}\big)^4\notag \\
&\Leftrightarrow&
(a^4+b^4)^3+a^45^{8k}=(c^4+d^4)^3+c^45^{8k},\label{injective025}
\end{eqnarray}
hence
\begin{equation}\label{injective04}
(a^4+b^4)^3\equiv (c^4+d^4)^3\pmod{5^{8k}}.
\end{equation}
Assume here first that $a^4+b^4$ and $c^4+d^4$ are not divisible by $5$. 
Since $|{\mathbb Z}^*_{5^{8k}}|$
$=\varphi(5^{8k})=5^{8k}(1-\frac{1}{5})=5^{8k-1}\cdot 4$ and
$\gcd(\varphi(5^{8k}),3)=1$, (\ref{injective04}) implies
\begin{equation}\label{injective05}
a^4+b^4\equiv c^4+d^4\pmod{5^{8k}},
\end{equation}
which together with (\ref{injective03}) implies $a^4+b^4=c^4+d^4$. But then
(\ref{injective025}) clearly implies that $a^4=c^4$, and hence also
$b^4=d^4$. As $a$, $b$, $c$, $d\ge 0$, it follows that $(a,b)=(c,d)$.

The only remaining thing is to check the case where either $a^4+b^4$ or $c^4+d^4$ is divisible by $5$. They both cannot, since in the beginning of the proof we chose $k$ minimal, and hence one of the numbers, $a$, $b$, $c$, $d$ is not divisible by $5$, and because by Fermat's little theorem, $n^4\equiv 1\pmod{5}$ for any integer $n$ not divisible by $5$. Without loss of generality, we can assume that for example, $a$ is not divisible by $5$. Then $a^4+b^4$ is either $\equiv 1$ or $\equiv2\pmod 5$, depending on wheter $b$ is divisible by $5$, but if both $c$ and $d$ are divisible by $5$ gives a contradiction against (\ref{injective04}).
\end{proof}

\subsection{Undecidability of injectivity for rational QFA}

We now have an injective polynomial $f(x, y) = (x^4+y^4)^3+x^4$ with domain $\Lambda \times \Lambda$  and range the rational numbers, where $\Lambda=\{\frac{a}{5^k}\mid a,k\in\mathbb N_0,a< 5^k\}$. As mentioned previously, we may extend polynomial $f$ to domain $\Lambda^n$ while retaining injectivity by defining $f_2(x_1, x_2) = f(x_1, x_2)$ and $f_k(x_1, \ldots, x_k) = f(x_1, f_{k-1}(x_2, \ldots, x_k))$ for $k > 2$ inductively.

As in Theorem~\ref{basistheorem2}, let $\Sigma=\{x_1, x_2, \ldots, x_{n-2}\}$ and $\Delta=\{x_{n-1}, x_n\}$ be distinct alphabets and $h, g: \Sigma^* \to \Delta^*$ be an instance of the mixed modification of PCP, with $\Sigma_n = \Sigma \cup \Delta$. Since $\gamma:\Sigma_n^* \to \Q^{4 \times 4}$ is a monomorphism, then it is undecidable to determine if there is a matrix in the following semigroup with two different factorizations:
\[
\Gamma = \langle \{\gamma(x_j) \oplus \gamma(h(x_j)), \gamma(x_j) \oplus \gamma(g(x_j)) | 1 \leq j \leq n-2\} \rangle \subseteq \Lambda^{8 \times 8}
\]
This follows by the same reasoning as in the proof of Theorem~\ref{basistheorem2}.

Notice that by Lemma~\ref{uniquequat2}, each element $X$ of the generator of $\Gamma$  (and indeed any product of these matrices) is uniquely determined by the six elements $\overline{x} = (|X_{1, 1}|, |X_{1, 2}|, |X_{1, 3}|, |X_{5, 5}|, |X_{5, 6}|, |X_{5, 7}|)$ since $X$ is a direct sum of two matrices from $\Q^{4 \times 4}$, each of which is determined by the absolute values of its top row, first three elements. This implies that $X$ is similarly uniquely determined by $\overline{x_2} = (X_{1, 1}^2, X_{1, 2}^2, X_{1, 3}^2, X_{5, 5}^2, X_{5, 6}^2, X_{5, 7}^2)$ which will be important later.

Our approach will be as follows. For each $X$ in the generator of $\Gamma$, we will derive a (much larger) matrix $\zeta_X$, as well as appropriately sized projection matrix $P$ and initial vector $u$, so that for any $Y = X_{j_1} \cdots X_{j_\ell} \in \Gamma$ the following equation holds:
\[
|P\zeta_{Y} u|^2 = f_6(Y_{1, 1}^2, Y_{1, 2}^2, Y_{1, 3}^2, Y_{5, 5}^2, Y_{5, 6}^2, Y_{5, 7}^2),
\]
where $\zeta_Y = \zeta_{X_{j_1}} \cdots \zeta_{X_{j_\ell}}$. This value is thus unique for each matrix $\zeta_{Y}$. 
In order to achieve our aim, we will use the Kronecker product, direct sums and require the following well known theorem of Lagrange. The reason for this is that some coefficients in $f_6$ are not perfect squares and in our encoding we would, without the use of Lagrange's theorem, require the use of radicals (which in this section we are explicitly trying to avoid).

\begin{theorem}[Lagrange]\label{lagrangethm}
Given a natural number $k$, there exist integers $a_1, a_2, a_3$ and $a_4$ such that
$$
k = a_1^2 + a_2^2 + a_3^2 + a_4^2
$$
\end{theorem}

Consider the injective polynomial $f_6$ of degree $d$ ($d = 4^{3^5}$) with integral coefficients. Let $\overline{x} = (|X_{1, 1}|, |X_{1, 2}|, |X_{1, 3}|, |X_{5, 5}|, |X_{5, 6}|, |X_{5, 7}|) \in \Q^6$, then we may write
$$
f_6(\overline{x}) = \sum_{i = 1}^{d} T_i(\overline{x})
$$ 
where $T_i(\overline{x})$ denotes the sum of terms of $f_6(\overline{x})$ of degree $1 \leq i \leq d$ (some $T_i$ may thus be zero). We may define $T_{i,j}(\overline{x})$ to be the $j$'th term of $T_i(\overline{x})$ (where $T_i(\overline{x})$ is written as a sum of arbitrary but fixed order of terms $T_{i,j}(\overline{a})$). Let $t(i) \geq 0$ denote the number of terms of $f_6$ of degree $i$ and thus 
$$
f_6(\overline{x}) = \sum_{i = 1}^{d} \sum_{j = 1}^{t(i)} T_{i,j}(\overline{x})
$$ 

For each monomial $T_{i,j}(\overline{x})$, we may define that

$$
T_{i,j}(\overline{x}) = c_{i,j} R_{i,j}(\overline{x}),
$$
with $c_{i,j} \in \mathbb{N}$ a coefficient and $R_{i,j}(\overline{x}) = \prod_{m = 1}^{i}a_{i,j,m}$ a monomial with $a_{i,j,m} \in \{|X_{1, 1}|, |X_{1, 2}|, |X_{1, 3}|, |X_{5, 5}|, |X_{5, 6}|, |X_{5, 7}|\}$. For each coefficient $c_{i,j}$ we may write $c_{i,j} = d^2_{i,j,1} + d^2_{i,j,2} + d^2_{i,j,3} + d^2_{i,j,4}$ where each $d^2_{i,j,k} \in \N$ by Lagrange's theorem (Theorem~\ref{lagrangethm}) and thus
\begin{eqnarray}
f_6(\overline{x}) & = & \sum_{i = 1}^{d} \sum_{j = 1}^{t(i)} \sum_{k=1}^{4} d_{i,j,k}^2 R_{i,j}(\overline{x}) \label{theeqn} \\
& = & \sum_{i = 1}^{d} \sum_{j = 1}^{t(i)} \sum_{k=1}^{4} d_{i,j,k}^2 \prod_{m = 1}^{i}a_{i,j,m} \qquad | a_{i,j,m} \in \{|X_{1, 1}|, |X_{1, 2}|, |X_{1, 3}|, |X_{5, 5}|, |X_{5, 6}|, |X_{5, 7}|\} \nonumber
\end{eqnarray}

We are now in a position to describe our approach. Let $c_{i,j} R_{i,j}(\overline{x})$ be one fixed term of $f_6(\ov{x})$ with $1 \leq i \leq d$ and $j \leq t(i)$ with $c_{i,j} = \sum_{1 \leq k \leq 4}d_{i,j,k}^2$ for $d_{i,j,k} \in \N$, which are guaranteed to exist by Lagrange's theorem.  For each such term, we now find a homomorphism $\zeta'_{i,j}:\mathbb{Q}^{8 \times 8} \to \mathbb{Q}^{8^i \times 8^i}$ such that some particular cell (say row $s$ and column $r$ with $1 \leq s, r \leq 8^i$) then $\zeta'_{i,j}(X)_{s, r} = R_{i,j}(\ov{x})$. 

Let us describe how to determine $\zeta'_{i,j}(X)$ for each $X$ in the generator of $\Gamma$ with $\overline{x} = (|X_{1, 1}|, |X_{1, 2}|, |X_{1, 3}|, |X_{5, 5}|, |X_{5, 6}|, |X_{5, 7}|) \in \Q^6$. We wish to enforce that $\zeta'_{i,j}(X)_{s, r} = R_{i,j}(\ov{x})$. Let $\zeta'_{i,j}(X) = X^{\otimes i}  \in \Q^{8^i \times 8^i}$. We note that $X^{\otimes i}$ contains elements which are all possible products of exactly $i$ elements of matrix $X$. Therefore there exist a row $s$ and column $r$ such that $X^{\otimes i}_{s,r} = R_{i,j}(\ov{x})$.

It is then easy to see that for vector $u'_{i,j,k} = d_{i,j,k}\cdot e_{r} \in \Q^{8^i}$ and matrix $P'_{i,j} = e_se_s^T  \in \Q^{8^i \times 8^i}$ (with $e_r$ the basis vector with a $1$ at position $r$ and zero elsewhere): 
$$
P'_{i,j}\zeta'_{i,j}(X) u'_{i,j,k} = d_{i,j,k} R_{i,j}(\ov{x})
$$
Now, defining $P_{i,j} = \bigoplus_{k=1}^{4} P'_{i,j}  \in \Q^{4*8^{i} \times 4*8^{i}}$, $u_{i,j} = \bigoplus_{k=1}^{4} u'_{i,j,k}  \in \Q^{4*8^{i}}$ and $\zeta_{i,j}(X) = \bigoplus_{k=1}^{4} \zeta_{i,j}'(X)  \in \Q^{4*8^{i} \times 4*8^{i}}$ we see that:
\begin{eqnarray*}
\left|\left| P_{i,j} \zeta_{i,j}(X) u_{i,j} \right|\right|^2 & = & \left|\left| \bigoplus_{k = 1}^{4} P'_{i,j} \zeta'_{i,j}(X) u'_{i,j,k} \right|\right|^2 \\
& = & \left( \sqrt{\sum_{k=1}^{4} d^2_{i,j,k} R_{i,j}(\ov{x})^2}\right)^2 = \sum_{k=1}^{4} d^2_{i,j,k} R_{i,j}(\ov{x})^2 = c_{i,j} R_{i,j}(\ov{x_2})
\end{eqnarray*}

Let us make a few observations here. Recall that $c_{i,j} = d^2_{i,j,1} + d^2_{i,j,2} + d^2_{i,j,3} + d^2_{i,j,4}$ by definition. Notice also that $R_{i,j}(\ov{x})^2 = R_{i,j}(\ov{x_2})$, i.e., 
\[R_{i,j}(|X_{1, 1}|, |X_{1, 2}|, |X_{1, 3}|, |X_{5, 5}|, |X_{5, 6}|, |X_{5, 7}|)^2 =  R_{i,j}(X_{1, 1}^2, X_{1, 2}^2, X_{1, 3}^2, X_{5, 5}^2, X_{5, 6}^2, X_{5, 7}^2).\] 
Finally, note that each $P'_{i,j} \zeta'_{i,j}(X) u'_{i,j,k}$ is a vector with exactly one nonzero element. Indeed, 
$$
P'_{i,j} \zeta'_{i,j}(X) u'_{i,j,k} = e_se_s^T \zeta'_{i,j}(X) d_{i,j,k} e_r = d_{i,j,k}\cdot \zeta'_{i,j}(X)_{s, r} \cdot e_s = d_{i,j,k} R(\ov{x}) \cdot e_s
$$

The described approach allows us to define matrices and vectors so that $\left|\left| P_{i,j} \zeta_{i,j}(X) u_{i,j} \right|\right|^2 = c_{i,j} R_{i,j}(\ov{x_2})$, although we note that $u$ is not a unit length vector and thus this formalism does not yet correspond to a QFA.

The above approach can be used for each term individually by using the direct sum. By using Eqn.~\ref{theeqn} we form the following vector and matrices:
$$
P' = \bigoplus_{i=1}^{d} \bigoplus_{j=1}^{t(i)} \bigoplus_{k=1}^{4} P_{i,j}, \quad \zeta' = \bigoplus_{i=1}^{d} \bigoplus_{j=1}^{t(i)} \bigoplus_{k=1}^{4} \zeta_{i,j} = \bigoplus_{i=1}^{d} \bigoplus_{j=1}^{t(i)} \bigoplus_{k=1}^{4} \bigotimes_{m=1}^{i} X, \quad u' = \bigoplus_{i=1}^{d} \bigoplus_{j=1}^{t(i)} \bigoplus_{k=1}^{4} u_{i,j,k}  
$$

By the mixed product properties of the Kronecker product (see Lemma~\ref{kronprop}), this structure is retained under matrix products. Notice that 
$$
||P'\zeta'u'||^2 =  \sum_{i=1}^{d} \sum_{j=1}^{t(i)} c_{i,j} R_{i,j}(\ov{x_2}) = f_6(\ov{x_2})
$$
as required. It remains to show how to make $u'$ a unit vector; note that $P'$ is a projection matrix (since it is the direct sum of projection matrices) and $\zeta'$ is a unitary matrix (as the direct sum and product of unitary matrices). Recall that 
$$
u' = \bigoplus_{i=1}^{d} \bigoplus_{j=1}^{t(i)} \bigoplus_{k=1}^{4} u_{i,j,k} = \bigoplus_{i=1}^{d} \bigoplus_{j=1}^{t(i)} \bigoplus_{k=1}^{4} d_{i,j,k} \cdot e_{r_{i,j,k}},
$$
where $e_{r_{i,j,k}}$ is of a fixed dimension $y$ and $r_{i,j,k} \leq y$ is some integer. We then see that 
\[
|u'| = \sqrt{\sum_{i=1}^d\sum_{j=1}^{t(i)}\sum_{k=1}^4 d_{i,j,k}^2}
\]
We notice then that $|u'|$ may not be rational unless the inner summations produce a perfect square. 
We may thus find $\delta \in \N$ such that $\delta + \sum_{i=1}^d\sum_{j=1}^{t(i)}\sum_{k=1}^4 d_{i,j,k}^2$ is a perfect square. Using Lagrange's theorem, we write $\delta = \delta_1^2 + \delta_2^2 + \delta_3^2 + \delta_4^2$ where $\delta_i \in \N$. Then we define $u'' = u' \oplus \delta_1 \oplus \delta_2 \oplus \delta_3 \oplus \delta_4$ so that 
\[
|u''| = \sqrt{\delta_1^2 + \delta_2^2 + \delta_3^2 + \delta_4^2 + \sum_{i=1}^d\sum_{j=1}^{t(i)}\sum_{k=1}^4 d_{i,j,k}^2},
\]
which is now a rational since $|u''|$ is a perfect square and thus let $u = \frac{u''}{|u''|}$ be the rational initial vector. We define each $\zeta = \zeta' \oplus I_4$ where $I_4$ is the $4 \times 4$ identity matrix and thus each $\zeta$ is still unitary since it is the direct sum of unitary matrices. We therefore have one such $\zeta$ for each $X \in \Gamma$, denoted $\zeta_{1}, \ldots, \zeta_{2n-4}$. We define $P = P' \oplus {\bf 0}_4$ where ${\bf 0}_4$ is the $4 \times 4$ zero matrix and thus $P$ is still a projection matrix. We now have a quantum finite automaton defined by $(P, \{\zeta_{\ell}|1 \leq \ell \leq 2n-4\}, u)$ such that 
\[
|PYu|^2 = f_6(Y_{1,1}^2, Y_{1,2}^2, Y_{1,3}^2, Y_{5,5}^2, Y_{5,6}^2, Y_{5,7}^2),
\]
for any $Y \in \langle \{\zeta_{\ell}|1 \leq \ell \leq 2n-4\} \rangle$ and since $f_6$ is an injective polynomial over $\Lambda^6$, then $\mathcal{Q}$ is not injective if and only if the instance of mixed modification of PCP $h, g$ has a solution.

We may note that $\mathcal{Q}$ has a finite number of states although each monomial corresponds to a matrix of dimension no more than $8^{4^{3^5}}$. 

\section{Conclusion}

We showed that determining if a measure once quantum finite automaton (QFA) is injective is undecidable when the automaton is defined over algebraic reals when the QFA is of dimension $8$ and when the QFA is defined over rationals, although with a huge (of the order $8^{4^{3^5}}$) number of states. Both of these results are derived for a fixed size input alphabet. This dimension would be reduced significantly if an injective polynomial of lower degree for $n$-tuples of rationals can be found, although this is a much more general problem. We derive restricted injective polynomials $f_n$ over a domain $\Lambda^n$ where $\Lambda = \{\frac{a}{5^k}\mid a,k\in\mathbb N_0,a< 5^k\}$ which are sufficient for our purposes. We also used some properties of quaternions. Firstly we use the fact that the semigroup $\mathcal{S}$ generated by $\{q_a, q_b\} \subseteq \mathbb{H}$ is free where $q_a = (\frac{3}{5}, \frac{4}{5}\mathbf{i}, 0, 0)$ and $q_b = (\frac{3}{5}, 0, \frac{4}{5}\mathbf{j}, 0)$, which follows by a result of Swierczowski. Secondly we derive the property that a quaternion $q = (q_1, q_2{\bf i}, q_3{\bf j}, q_4{\bf k}) \in \mathcal{S}$ with $q_1, q_2, q_3, q_4 \in \mathbb{Q}$ is uniquely determined by $(|q_1|, |q_2|, |q_3|)$.

We note that in \cite{BCJ16} the ambiguity and freeness problems for weighted finite automata and probabilistic finite automata were shown to be undecidable even when the input words were restricted to come from a given \emph{letter-bounded language}, which is a restriction of bounded languages of the form $x_1^*x_2^* \cdots x_k^*$ where each $x_i$ is a single letter of the input alphabet. The undecidability result of \cite{BCJ16} used an encoding of Hilbert's tenth problem, which seems difficult to encode into unitary matrices and thus we pose the following open problem.

\begin{open}
Can the undecidability of the ambiguity and freeness problems for MO-QFA be shown when the input word is necessarily from a given letter-bounded language?
\end{open}

\noindent{\bf Acknowledgements.} We would like to thank the two reviewers for their careful checking of this manuscript and their helpful suggestions.

\bibliography{refs}

\providecommand{\BIBYu}{Yu}
\begin{thebibliography}{10}
\expandafter\ifx\csname url\endcsname\relax
  \def\url#1{\texttt{#1}}\fi
\expandafter\ifx\csname urlprefix\endcsname\relax\def\urlprefix{URL }\fi
\expandafter\ifx\csname href\endcsname\relax
  \def\href#1#2{#2} \def\path#1{#1}\fi

\bibitem{MC}
C.~Moore, J.~P. Crutchfield, Quantum automata and quantum grammars, Theoretical
  Computer Science 237~(1-2) (2000) 275--306.

\bibitem{Brodsky2002}
A.~Brodsky, N.~Pippenger, Characterizations of 1-way quantum finite automata,
  SIAM Journal on Computing 31 (2002) 1456--1478.

\bibitem{Paz}
A.~Paz, Introduction to Probabilistic Automata, Academic Press, 1971.

\bibitem{BJK05}
V.~Blondel, E.~Jeandel, P.~Koiran, N.~Portier, Decidable and undecidable
  problems about quantum automata, SIAM Journal on Computing 34:6 (2005)
  1464--1473.

\bibitem{Hir}
M.~Hirvensalo, Improved undecidability results on the emptiness problem of
  probabilistic and quantum cut-point languages, SOFSEM 2007: Theory and
  Practice of Computer Science, Lecture Notes in Computer Science 4362 (2007)
  309--319.

\bibitem{Paterson}
M.~S. Paterson, Unsolvability in $3\times 3$ matrices, Studies in Applied
  Mathematics 49~(1) (1970) 105--107.

\bibitem{BP10}
P.~C. Bell, I.~Potapov, On the undecidability of the identity correspondence
  problem and its applications for word and matrix semigroups, International
  Journal of Foundations of Computer Science 21~(6) (2010) 963--978.

\bibitem{KBS91}
D.~A. Klarner, J.-C. Birget, W.~Satterfield, On the undecidability of the
  freeness of integer matrix semigroups, International Journal of Algebra and
  Computation 1 (2) (1991) 223--226.

\bibitem{CHK99}
J.~Cassaigne, T.~Harju, J.~Karhum\"aki, On the undecidability of freeness of
  matrix semigroups, International Journal of Algebra and Computation 9~(3-4)
  (1999) 295--305.

\bibitem{CN12}
J.~Cassaigne, F.~Nicolas, On the decidability of semigroup freeness, RAIRO -
  Theoretical Informatics and Applications 46~(3) (2012) 355--399.

\bibitem{CH14}
E.~Charlier, J.~Honkala, The freeness problem over matrix semigroups and
  bounded languages, Information and Computation 237 (2014) 243--256.

\bibitem{CK05}
C.~Choffrut, J.~Karhum\"aki, Some decision problems on integer matrices,
  Informatics and Applications 39 (2005) 125--131.

\bibitem{BP08quat}
P.~C. Bell, I.~Potapov, Reachability problems in quaternion matrix and rotation
  semigroups, Information and Computation 206~(11) (2008) 1353--1361.

\bibitem{BPSofsem08}
P.~C. Bell, I.~Potapov, Periodic and infinite traces in matrix semigroups,
  Current Trends in Theory and Practice of Computer Science (SOFSEM) LNCS 4910
  (2008) 148--161.

\bibitem{KP18}
S.-K. Ko, I.~Potapov, Vector ambiguity and freeness problems in {SL}($2$,
  $\mathbb{Z}$), Fandumenta Informaticae 162~(2-3) (2018) 161--182.

\bibitem{COSW}
T.~Colcombet, J.~Ouaknine, P.~Semukhin, J.~Worrell, On reachability problems
  for low dimensional matrix semigroups, in: ArXiV Manuscript (to appear
  ICALP'19), Vol. arXiv:1902.09597, 2019, pp. 1--15.

\bibitem{BCJ16}
P.~C. Bell, S.~Chen, L.~M. Jackson, Scalar ambiguity and freeness in matrix
  semigroups over bounded languages, in: Language and Automata Theory and
  Applications, Vol. LNCS 9618, 2016, pp. 493--505.

\bibitem{KS86}
W.~Kuich, A.~Salomaa, Semirings, Automata, Languages, Vol.~5, Springer, 1986.

\bibitem{Ho98}
J.~Honkala, Decision problems concerning thinness and slenderness of formal
  languages, in: Acta Informatica, Vol.~35, 1998, pp. 625--636.

\bibitem{BC03}
V.~Blondel, V.~Canterini, Undecidable problems for probabilistic automata of
  fixed dimension, Theory of Computing Systems 36 (2003) 231--245.

\bibitem{BHH13}
P.~C. Bell, V.~Halava, M.~Hirvensalo, Decision problems for probabilistic
  finite automata on bounded languages, Fundamenta Informaticae 123~(1) (2012)
  1--14.

\bibitem{BMT77}
A.~Bertoni, G.~Mauri, M.~Torelli, Some recursively unsolvable problems relating
  to isolated cutpoints in probabilistic automata, in: Automata, Languages and
  Programming, Vol.~52 of LNCS, 1977, pp. 87--94.

\bibitem{BH19}
P.~C. Bell, M.~Hirvensalo, Acceptance ambiguity for quantum automata, in:
  Mathematical Foundations of Computer Science, no.~70 in MFCS'19, 2019, pp.
  1--14.

\bibitem{Le04}
E.~Lengyel, Mathematics for 3D Game Programming \& Computer Graphics, Charles
  River Media, 2004.

\bibitem{HJ91}
R.~A. Horn, C.~R. Johnson, Topics in matrix analysis, Cambridge University
  Press, 1991.

\bibitem{CKH96}
J.~Cassaigne, J.~Karhum\"aki, T.~Harju, On the decidability of the freeness of
  matrix semigroups, International Journal of Algebra and Computation 9~(3-4)
  (1999) 295--305.

\bibitem{Cl80}
V.~Claus, Some remarks on {PCP}($k$) and related problems, Bulletin of the
  EATCS 12 (1980) 54--61.

\bibitem{HHH06}
V.~Halava, T.~Harju, M.~Hirvensalo, Undecidability bounds for integer matrices
  using {C}laus instances, International Journal of Foundations of Computer
  Science (IJFCS) 18,5 (2007) 931--948.

\bibitem{Sw94}
S.~Swierczkowski, A class of free rotation groups, Indag. Math. 5~(2) (1994)
  221--226.

\bibitem{Lang2002}
S.~Lang, Algebra, Springer, 2002.

\bibitem{FueterPolya}
R.~Fueter, G.~P\'olya, Rationale abz\"ahlung der gitterpunkte,
  Vierteljahrsschrift der Naturf. Gesellschaft in Z\"urich 58~(1) (1923)
  380--386.

\bibitem{Vsemirnov}
M.~A. Vsemirnov, Two elementary proofs of the fueter-p\'olya theorem on
  matching polynomials, Algebra i Analiz 13~(5) (2001) 1--15.

\bibitem{Adriaans}
P.~W. Adriaans, A simple information theoretical proof of the fueter-pólya
  conjecture (2018).

\end{thebibliography}

\end{document}